\documentclass[%
reprint,
amsmath,
amssymb,
aps,
nofootinbib,
superscriptaddress,
floatfix]{revtex4-1}

\usepackage{graphicx}
\usepackage{color}
\usepackage{ulem} 
\usepackage{braket}
\usepackage{comment}

\DeclareGraphicsExtensions{.eps}

\def\punkt{\;\; .}
\def\komma{\;\; ,}

\def\mat#1{\mathbf{#1}}
\def\w{\omega}

\def\e{\epsilon}

\def\trb#1{{\rm Tr_R}\left[ #1\right]}
\def\Tr#1{\textrm{Tr}\left[#1\right]}
\def\non{\nonumber\\ }

\def\myeqref#1{Eq.~\eqref{#1}}

 \def\correl#1#2{ G_{ #1 ; #2}}
 
\begin{document}

\title{Restoring the continuum limit in the time-dependent numerical renormalization group approach}

\author{Jan B\"oker}
\author{Frithjof B. Anders}
\affiliation{Lehrstuhl f\"ur Theoretische Physik II, Technische Universit\"at Dortmund
Otto-Hahn-Str. 4, 44227 Dortmund,
Germany}

\date{\today}

\begin{abstract}
The continuous coupling function in quantum impurity problems is exactly partitioned into a part represented by a finite size Wilson chain and a part represented by a set of additional reservoirs, each coupled to one Wilson chain site. These additional reservoirs represent high-energy modes of the environment neglected by the numerical renormalization group and are required
to restore the continuum limit of the original problem. We present a hybrid time-dependent numerical renormalization group approach which combines an accurate numerical renormalization group treatment of the non-equilibrium dynamics on the finite size Wilson chain with a Bloch-Redfield formalism to include the effect of these additional reservoirs. Our approach overcomes the intrinsic shortcoming of the time-dependent numerical renormalization group approach induced by the bath discretization with a Wilson parameter $\Lambda > 1$. We analytically prove that for a system with a single chemical potential, the thermal equilibrium reduced density operator is the steady-state solution of the Bloch-Redfield master equation. For the numerical solution of this master equation a Lanczos method is employed which couples all energy shells of the numerical renormalization group. 
The presented hybrid approach is applied to the real-time dynamics in correlated fermionic quantum-impurity systems. An analytical solution of the resonant-level  model  serves as a benchmark for the accuracy of the method which is then applied to non-trivial models, such as the interacting resonant-level model and the single impurity Anderson model.

\end{abstract}
\maketitle

\section{Introduction}

Quantum impurity systems (QIS) have been of  increasing interest in the last two decades due to the
advent of single-electron transistors \cite{KastnerSET1992} and the observation of the Kondo effect in
nano devices \cite{NatureGoldhaberGordon1998,goldhaberSET98,Kouwenhoven2000} as well as in adatoms
 \cite{Manoharan2000,AgamSchiller2001}  and molecules \cite{TemirovLassieAndersTautz2008} on surfaces. Charge and spin dynamics of molecules on surfaces \cite{galperinNitzanRatner2006,MolecularElectronicsReview2009rohtua} including inelastic processes \cite{LorentePersson2000,REED2008,EickhoffSTM2020} as well as local moment formations and quantum phase transitions in the vicinity of graphene vacancies \cite{Pareira2007,Cazalilla2012,MayGraphen2018,AndreiGraphen2018} are only a few examples of many such different realizations. QIS are also of fundamental importance as a part of the dynamical mean field theory  \cite{Kuramoto85,Georges96} where a correlated lattice problem is mapped onto an effective QIS \cite{KotliarVollhardt2004} augmented by a self-consistence condition.

On the route to functional nano devices, the real-time dynamics of local charge \cite{Elzerman04} or spin degrees of freedom \cite{HansonSpinQdotsRMP2007} sparked  the theoretical interest in the non-equilibrium dynamics of
observables in such systems \cite{Leggett1987,review1DQT}. Charge transfer and energy-transfer dynamics in molecular systems have also been investigated for more than two decades \cite{MayKuehn2000}. 

The theoretical approaches addressing the non-equilibrium dynamics can be divided into three categories. The first class of approaches relies on  partitioning the full continuum Hamiltonian into an exactly solvable part and a residue treated as a perturbation. Amongst those are the Keldysh diagrammatic approaches \cite{Keldysh65,KadanoffBaym62,LangrethWilkins1972} to quantum impurity problems \cite{JauhoWingreenMeir1994,NordlanderEtAl1999} as well as more advanced functional renormalization group \cite{FRG,RT-RG-IRLM-2}, real-time renormalization group \cite{Schoeller2009a} 
and flow equation methods \cite{Wegner1994,Kehrein2005}. The extension of diagrammatic quantum Monte Carlo methods \cite{GulletAl2011} to the real-time dynamics suffers from a sign problem \cite{MuehlbacherRabani2008,SchmidtWerner2008,Schiro2010} which has been tamed by the worm inch algorithm \cite{CohenMillisPRL2015} only recently.
The second class of approaches replaces the closed continuum problem by a finite size representation  of relevant impurity degrees of freedom subject to a Lindblad or Bloch-Redfield master equation \cite{CarmichaelQuantumOpticsI,MayKuehn2000}. Such approaches have been proposed for systems that are coupled only weakly to their environment but also have been extended to more complex QIS \cite{NussArrigoni2015,DoraArrigioni2015}  targeting quantum transport problems out of equilibrium. The latter extension uses the Lindblad decay rates as fitting parameters to reproduce the continuum limit of the non-interacting part of the original problem as accurately as possible.
The third class of methods performs a mapping of the original continuum problem onto a discretized representation which is then treated by exact diagonalization \cite{Kuijaars2000,SaadSparseLinearSystems2003}, pure state propagation \cite{TalEzer-Kosloff-84,Kosloff-94,Fehske-RMP2006,Steinigeweg2014,HackmannAnders2014}
by the time-dependent numerical renormalization group (TD-NRG) \cite{AndersSchiller2005,AndersSchiller2006,AndersSSnrg2008,NghiemCosti2014,CostiGF2017} or the time-dependent density matrix renormalization group (TD-DMRG) approach \cite{DaleyKollathSchollwoeckVidal2004,Schollwoeck-2005,Schollwoeck2011}.

In this paper, we propose a  hybrid TD-NRG approach that combines the virtue of the NRG \cite{AndersSchiller2005,AndersSchiller2006,BullaCostiPruschke2008} encoding an accurate representation of equilibrium fixed points with a Bloch-Redfield master equation approach \cite{MayKuehn2000} in order to restore the original continuum problem. In the previous hybrid TD-NRG algorithms  different numerical methods 
(TD-NRG and Chebyshev polynomials \cite{EidelsteinGuettgeSchillerAnders2012}
or TD-NRG and TD-DMRG \cite{GuettgeAndersSchiller2013})
were combined but still operated on a finite one-dimensional chain representation of the Hamiltonian and did not solve the fundamental limitation of all finite size representations: true relaxation and thermalization. In a chain representation of the problem, the continuity equations derived from charge conservation lead to backreflexions within the Wilson chain  \cite{EidelsteinGuettgeSchillerAnders2012} or at the end of a tight-binding chain \cite{Schmitteckert2006}.

We make use of the exact decomposition of the bath continuum into the Wilson chain and augmented reservoirs attached to each chain site.  We adopt the proposal \cite{SBMopenchain2017} made  in the context of the spin boson model \cite{Leggett1987} to fermionic baths. 
In the pervious work \cite{SBMopenchain2017}  only corrections to the Wilson chain parameters obtained from the real part of the bosonic reservoir coupling function were included in the calculations for the spin boson model \cite{BullaBoson2003}. Here, we link the Bloch-Redfield tensor \cite{MayKuehn2000} to the previously neglected imaginary parts of the  fermionic reservoir correlation functions: these tensor elements govern the real-time dynamics of the reduced density matrix by connecting NRG eigenstates on different Wilson shells \cite{AndersSchiller2005,AndersSchiller2006,BullaCostiPruschke2008} or NRG iterations.
In our algorithm the static reduced density matrix in the TD-NRG \cite{AndersSchiller2005,AndersSchiller2006} is  replaced by a time-dependent version and its dynamics is generated by the previously neglected reservoirs. Our approach conserves the trace of the density matrix at any time and approaches thermal equilibrium as the steady-state solution for any  Bloch-Redfield tensor that fulfills the generic detailed balance condition. Therefore, our approach corrects the drawback of all finite size real-time methods, namely that a true stationary steady state can only be reached in the limit of an infinite system size that is not accessible for such methods.

The paper is organized as follows. In Sec.\ \ref{Sec:discretisation-and-continuum} we introduce the generic quantum
impurity model and derive the exact hybrid Wilson-chain continuum representation of the original coupling function in Sec.\ \ref{sec:chain-plus-single-reservoir}. In Sec. \ref{sec:reservoir}, we show that the resulting reservoir coupling functions approach  two alternating fixed points: one for the even chain sites and one for the odd chain sites that is typical for fermionic baths \cite{Wilson75,KrishWilWilson80a}. The proposed hybrid approach is presented in Sec.\ \ref{sec:non-equilibrium}. After a short review of the TD-NRG to introduce the notation, 
we derive the effect of the additional reservoirs up to second order in the fermionic coupling functions in Sec.\ \ref{sec:bloch-redfield} which are used in Sec.\ \ref{sec:neq-reduced-DM} to obtain the non-equilibrium dynamics of the reduced density matrix providing the essential of the hybrid approach. Some technical details about the implementation are provided in Sec.\ \ref{sec:algorithm}. In Sec.\ \ref{sec:benchmark} we present the benchmark for our approach by demonstrating the excellent agreement between the predictions of the continuum hybrid TD-NRG approach and the exact analytic solution of the charge dynamics in the resonant-level model (RLM)  \cite{AndersSchiller2006}. The non-equilibrium dynamics of two correlated models, the interacting RLM \cite{VigmanFinkelstein78-1,VigmanFinkelstein78-2,Schlottmann1980} and the single impurity Anderson model  (SIAM) \cite{KrishWilWilson80a} are discussed, and the paper ends with a short summary. 

\section{Discretization and restoring of the continuum limit}
\label{Sec:discretisation-and-continuum}

\subsection{Introduction to quantum impurity models}

Quantum impurity models (QIS) describe the coupling of a strongly interacting
quantum impurity $H_{\rm imp}$ with  non-interacting baths $H_{\rm bath}$ comprising
either conduction bands \cite{Wilson75,KrishWilWilson80a} or a bosonic
environment \cite{Leggett1987}:
\begin{eqnarray}
H &=H_{\rm imp} +H_{\rm bath} + H_{I} \punkt
\label{eq:first_H}
\end{eqnarray}
The  term $H_{I}$ describes the interaction between the two subsystems. 
$H_{\rm bath}$ models $M$ different 
non-interacting and continuous fermionic baths
\begin{eqnarray}
\label{eqn:bath}
H_{\rm bath} &=& \sum_{\nu=1}^M \sum_{k} \e_{k\nu} c^\dagger_{k\nu}c_{k\nu}^{ }
\end{eqnarray}
with the flavors $\nu$.  
$c^\dagger_{k\nu}$ creates a bath electron  of flavor $\nu$ with
the energy $\e_{k\nu}$.  $\nu$ might label the spin $\sigma$ or
the channel $\alpha$ in multi-band models. We focus on a coupling $H_{I}$ between
the two subsystems described by  a single particle hybridization
\begin{eqnarray}
\label{eqn:Hhyp}
H_I &=& \sum_{\nu=1}^{M}V_\nu  \left(c^\dagger_{0\nu} A_\nu^{ } + A^\dagger_{\nu} c_{0\nu}^{ }\right),
\end{eqnarray}
where $c_{0\nu}$ annihilates a local bath state of flavor $\nu$ defined as a linear combination of annihilators $ c_{k\nu}$ of bath modes with the eigenenergy $\e_{k\nu}$
 \begin{eqnarray}
\label{eqn:c-local-orig}
c_{0\nu} &=& \sum_{k} \lambda_{k\nu} c_{k\nu}
\end{eqnarray}
such that $c_{0\nu}$ fulfils canonical commutation relations.
$A^\dagger_\nu (A_\nu)$  accounts for the  linear combination of local orbital creation (annihilation) operators inducing
transitions in the impurity that change the particle number by one.
The coupling parameters $\lambda_{k\nu}$
contain the possible energy-dependent hybridization.

By integrating out the bath degrees of freedom in a path integral formulation of the partition function, 
it has been noted early on \cite{Wilson75,Leggett1987,BullaPruschkeHewson1997} that the
influence of the bath onto the local impurity dynamics 
is fully determined by the coupling function $\Delta_\nu(z)$ defined as
\begin{eqnarray}
\Delta_\nu(z) &=& V^2_{\nu} \sum_k \frac{\lambda^2_{k\nu}}{z-\e_{k\nu}}
\punkt
\end{eqnarray}
We will utilize the fact that different types of reservoirs \cite{BullaPruschkeHewson1997}
yield the same local dynamics as long as they provide the identical coupling functions $\Delta_\nu(z)$. The spectral function 
\begin{eqnarray}
\label{eq:gamma-nu}
\Gamma_\nu(\w)&=& \lim_{\delta \rightarrow 0^+} {\rm Im} \Delta_\nu(\w-i\delta)
\end{eqnarray}
determines the influence of the $\nu$-th bath onto  the local dynamics.
For nonsymmetric baths \cite{Leggett1987} the real part $\Re  e\Delta_\nu(\w)$ causes an 
additional energy renormalization of impurity eigenenergies. This energy renormalization strongly influences the dynamics close to a local quantum critical point \cite{vojtaErr2009,VojtaBullaGuettgeAnders2010,SBMopenchain2017}
in the case of bosonic baths but plays a less pronounced role in fermionic baths.

\subsection{Disretization of the continuum model}

\begin{figure}[t]
\includegraphics[width=0.47\textwidth]{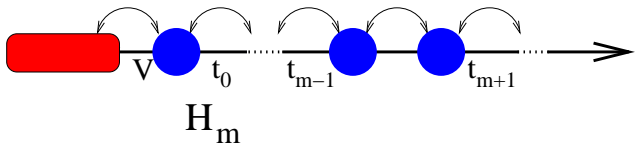}
\caption{The semi-infinite Wilson chain depicted up to the chain link $m$.}
\label{fig:1}
\end{figure}

The NRG \cite{Wilson75,BullaCostiPruschke2008} 
is one of the powerful methods developed 
to accurately solve QIS. Within this approach the bath continuum is discretized 
on a logarithmic mesh controlled by the parameter $\Lambda >
1$. The Hamiltonian is then mapped onto a semi-infinite chain 
\begin{eqnarray}
H^{\rm NRG} &=& \lim_{N\to\infty}H_{N}^{\rm NRG}\\
H_N^{\rm NRG} &=& H_{\rm imp}  +H_{I-C} + H_{\rm chain}(N)
\label{eqn:h-nrg-n}
\\
H_{\rm chain}(N) &=&
\sum_{m=0}^{N} \sum_{\nu=1}^M\e_{m\nu}^{ } f^\dagger_{m\nu} f_{m\nu}^{ }
\non
&&
+ \sum_{m=1}^{N}\sum_{\nu=1}^M t_{m-1 \nu}^{ }\left(f^\dagger_{m\nu} f_{m-1\nu}^{ } 
+f^\dagger_{m-1\nu} f_{m\nu}\right)
\non
H_{I-C} &=&\sum_{\nu=1}^M V_\nu^{ } \left(f^\dagger_{0\nu} A_\nu^{ } + A^\dagger_{\nu} f_{0\nu}^{ }\right)
\komma
\end{eqnarray}
whose chain topology is depicted in Fig.~\ref{fig:1}. The $m$-th chain site represents an exponentially 
decreasing energy scale 
$\w_m = D \Lambda^{-(m-1)/2} (1 +\Lambda^{-1})/2$, 
and the original Hamiltonian is only
restored \cite{Wilson75} in the limit $\Lambda\to 1^+$. The tight binding parameters $t_m$ 
also decrease exponentially, $t_m\propto \Lambda^{-m/2}$, which establishes  the
hierarchy of scales in the sequence of finite-size Hamiltonians $H^{\rm NRG}_m$. 
The bath asymmetry \cite{BullaCostiPruschke2008} mentioned above enters the single particle 
energies $\e_{m\nu}^{ }$ of each chain site.

This sequence of $H^{\rm NRG}_m$ is iteratively diagonalized, discarding the
high-energy states at each step to maintain a manageable number of
states.  Thereby,  the set of eigenstates of $H^{\rm NRG}_{m}$, $\{\ket{r,e;m}\}$, with the corresponding eigenenergies $E_r^{m}$ is partitioned into
a set of kept (k) states $S_k=\{\ket{k,e;m}\}$ and a set of  states $S_d=\{\ket{l,e;m}\}$ which will be discarded (d) in the next NRG 
iteration.  Since the iteration is stopped at a finite but arbitrary value $m=N$,
we have augmented the eigenstate $\ket{k}$ at iteration $m$ with the configuration $e$ of
the decoupled rest chain $m+1\to N$ to obtain a complete basis set - for details see
Refs.~\cite{AndersSchiller2005,AndersSchiller2006}.
The reduced basis set of $H^{\rm  NRG}_m$, $S_k$,  thus obtained is
expected to faithfully describe the spectrum of the full Hamiltonian
on the scale of $D_m$, corresponding \cite{Wilson75} to a temperature 
$T_m \sim D_m$ from which all thermodynamic expectation values are
calculated.  The NRG algorithm is stopped at chain length $N$ when the lowest temperature
of interest is reached.

In the present work, we will not discuss the explicit construction of such 
chains 
as a faithful representation of the original continuous baths and refer the reader to
the reviews \cite{Wilson75,SchollwoeckDMRG2005,BullaCostiPruschke2008} 
on this subject.
Here we assume that the NRG  framework has provided us already with all chain parameters
such as nearest neighbor hopping $t_{m\nu}^{ }$ and orbital energy $\e_{m\nu}^{ }$ of each
chain link to fully characterize any chain depicted in Fig.~\ref{fig:1}.

Independently of whether the NRG approach, exact diagonalization or  the  density 
matrix renormalization group (DMRG)~\cite{White92,SchollwoeckDMRG2005} 
is used to solve such a finite size representation of an interacting quantum impurity system,
these numerical approaches suffer from the same fundamental problem:
the finite size chain Hamiltonian does not contain any information on the life-time of excitations
and lacks the  mechanism for a locally excited system to relax into the true thermodynamic ground state.

This leads to two severe limitations when calculating the spectral functions
within the NRG: (i) details at high energies are lost by overbroadening 
($b_m\propto D_m$) even if the peak position and its spectral
weight are calculated correctly within the method,
and (ii) spectral information for frequencies below the smallest energy scale,
i.e.~$|\w|<D_N$, is absent which limits the accuracy of the NRG for calculating
transport 
properties \cite{GrenzebachAndersCzychollPruschke2006,GrenzebachAndersCzychollPruschke2008,BullaCostiPruschke2008}.

\subsection{Restoring the continuum limit}

\label{sec:chain-plus-single-reservoir}

\begin{figure}[tbp]

\includegraphics[width=0.45\textwidth=25mm]{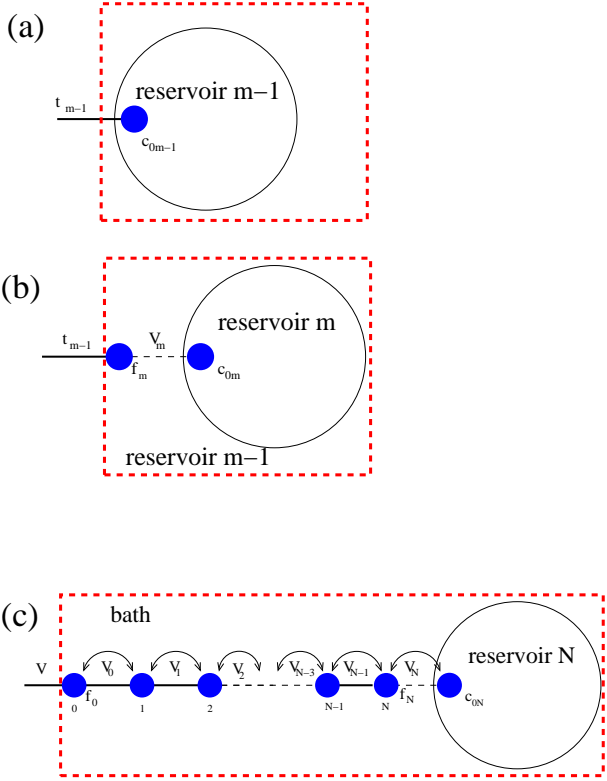}

\caption{The reservoir continuum $m-1$ (a) is recursively replaced by a single chain site $f_m$ 
coupled to a new reservoir degree of freedom $c_{0m}$ by a chain link matrix element $V_m$ , shown in (b), to
obtain a continuous fraction representation of the original bath by a (c) finite size tight-binding chain with the continuous reservoir coupled to the end of the chain as used in DMRG calculations \cite{KarskiRaasUhrig2008}.}
\label{fig:reservoir-iteration}

\end{figure}

We adapt   the approach \cite{SBMopenchain2017} introduced in the context
of the spin boson model \cite{Leggett1987} to fermionic baths
to reconstruct the correct hybridization function $\Delta_\nu(z)$ for a given Wilson chain.
We will drop the flavor index $\nu$ and restrict ourselves 
to a single flavor  for simplicity. We will restore the flavor index of the bath modes
at the end of this section.

Since the influence of the continuous bath onto
the  local dynamics of the quantum impurity is fully determined by the  function $\Delta(z)$, 
the bath Hamiltonian $\tilde H_{\rm bath}(1)$  defined as
\begin{eqnarray}
\label{eqn:chain-reservoir-step-1}
\tilde H_{\rm bath}(1) &=& \e_{0} f^\dagger_0 f_0^{ } + \sum_{k} \e_{k0}^{ }c^\dagger_{k0}c_{k0}^{ }
\non
&&
+ V_0\left(f^\dagger_0 c_{00}^{ } +c^\dagger_{00} f_0^{ }\right)
\end{eqnarray}
yields the same local dynamics 
as the original $H_{\rm bath}$  if  the Green function (GF) of the original bath $\correl{c_{0}^{ }}{c^\dagger_0}(z)$ is
identical to the GF  $\correl{f_{0}^{ }}{f^\dagger_0}(z)$
\begin{eqnarray}
\label{eqn:inital-gamma}
\Delta(z) &=& V^2 \correl{c_{0}^{ }}{c^\dagger_0}(z) = V^2 \correl{f_{0}^{ }}{f^\dagger_0}(z)
\komma
\end{eqnarray}
 and the hybridization in \eqref{eqn:Hhyp} is replaced by
\begin{eqnarray}
\label{eq:neq_H_I}
H_I &=& V\left(f^\dagger_{0} A + A^\dagger f_{0}^{ }\right)
\punkt
\end{eqnarray}

The index $1$ in $\tilde H_{\rm bath}(1)$ indicates that $H_{\rm bath}$ has been replaced by a new bath coupled to a single
auxiliary orbital. This new degree of freedom, $f_0$, will become the first site of the chain representation of the
bath continuum which we will construct in the following.
Analog to Eq.~(\ref{eqn:c-local-orig}), we have defined the new operator $c_{00}$ 
of the new reservoir $0$
\begin{eqnarray}
\label{eqn:c-00}
c_{00}^{ } &= & \sum_k \lambda_{k0}^{ } c_{k0}^{ }
\quad , \quad
\sum_k \lambda_{k0}^2 =1
\end{eqnarray}
as a linear combination of its reservoir modes.

The bath Hamiltonian (\ref{eqn:chain-reservoir-step-1}) describes a resonant level model whose
GF $\correl{f_{0}^{ }}{f^\dagger_0}(z)$ is given by
\begin{eqnarray}
\correl{f_{0}^{ }}{f^\dagger_0}(z) &=& \frac{1}{z-\e_0 - V_0^2 \correl{c_{00}^{ }}{c^\dagger_{00}}(z)}
\punkt
\end{eqnarray}
The unknown reservoir coupling  function 
$\Delta_0(z)$, defined as
\begin{eqnarray}
\Delta_0(z)& \equiv&V^2_0\correl{c_{00}^{ }}{c^\dagger_{00}}(z) 
\komma
\end{eqnarray}
is simply related to $\Delta(z)$ via Eq.~(\ref{eqn:inital-gamma})
\begin{eqnarray}
\Delta_0(z)& =&  z -\e_0^{ } - \frac{1}{\correl{f_{0}^{ }}{f^\dagger_0}(z) }
\non
&=& z -\e_0^{ } - \frac{V^2}{\Delta(z)}
\punkt
\label{eqn:gamma-0}
\end{eqnarray}
Since the spectrum of $\correl{c_{00}}{c^\dagger_{00}}(z)$ must be normalized to unity, 
the coupling constant $V^2_0$ cannot be chosen freely in the model but is determined by the integral
\begin{eqnarray}
\label{eqn:v0-integral}
\pi V^2_0 &=& \int_{-\infty}^{\infty} d\w\, {\rm Im} \Delta_0(\w-i\delta)
\komma
\end{eqnarray}
where $\e_0^{ }$ is  given by the first momentum of the spectrum of $\Delta(z)$
\begin{eqnarray}
\e_0^{ } &=& \frac{1}{\pi V^2} \int_{-\infty}^{\infty} d\w\, \w \, {\rm Im} \Delta(\w-i\delta)
\punkt
\end{eqnarray}

Now we can apply the same arguments as above to the new reservoir $\Delta_0(z)$ and substitute 
it by another resonant level model comprising of the 
second chain site of a chain coupled to the new reservoir $1$. 
Recursively, we replace the previous reservoir $m-1$  at iteration $m$, shown in
Fig.~\ref{fig:reservoir-iteration}(a),  by an effective resonant level model 
involving a new reservoir $m$ as
depicted in  Fig.~\ref{fig:reservoir-iteration}(b). After $N+1$ such steps we
obtain a chain of length $N+1$ which is coupled to a single 
reservoir $N$ at the end, as plotted in Fig.~\ref{fig:reservoir-iteration}(c). 

The resulting chain parameters $\{ V_m\}$ and $\{\e_m\}$ represent 
a  continuous fraction expansion with a finite length which
has  been successfully used in DMRG calculations \cite{KarskiRaasUhrig2008}.
The proper continuum limit is restored by adding a single additional reservoir  coupled to the last chain site whose properties are uniquely determined by the original coupling function $\Delta(z)$.
The tight-binding parameters $V_m$, however,  always remain of the
order of the original band width $D$ for all $m$ in this procedure 
and Wilson chains with their refined built-in energy hierarchy cannot be generated this way.

\begin{figure}[tbp]

 \includegraphics[width=0.45\textwidth]{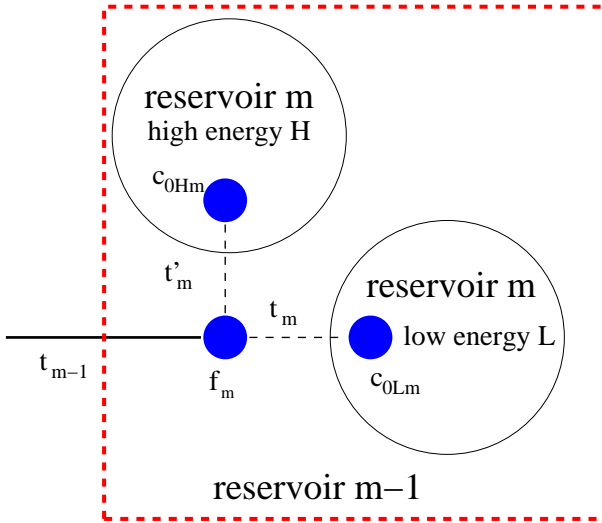}

\caption{
In the modified recursion for the Wilson chain, we divided the reservoir $m$ analog to  Fig.\ \ref{fig:reservoir-iteration}(b) into its high-energy and low-energy ($H$ and $L$ respectively) contribution which are tailored such that the low-energy part is coupled to $f_{m}$ with the matrix element $t_m<V_m$.
}
\label{fig:reservoir-iteration-NRG}

\end{figure}

In order to generate more general chains whose sites are coupled by 
arbitrary linking matrix elements $t_m$  ($t_m<V_m$) we need to supplement the algorithm with another step at each iteration. 
We assume that at some iteration $m$ the reservoir property is determined by a coupling
function $\tilde \Delta_{m-1}(z)$ such that the corresponding GF  is properly normalized
by the  coupling $t_{m-1}^2$  
\begin{eqnarray}
\label{eqn:define-tilde-delta-m}
\correl{c_{0m-1}^{ }}{c^\dagger_{0m-1}}(z) &=& \frac{1}{t_{m-1}^2} \tilde \Delta_{m-1}(z) 
\punkt
\end{eqnarray}
We will explicitly specify $\tilde \Delta_{m-1}(z)$ below by showing how it is determined
by the modified recursion. 
As before, we replace the reservoir $m-1$ by an additional chain site $m$  coupled to a new reservoir $m$ 
as  depicted
in Fig.~\ref{fig:reservoir-iteration}(b). The new reservoir coupling function is 
obtained by the same recursion
\begin{eqnarray}
\label{eqn:delta-m-recursion}
\Delta_{m}(z) &=& z -\tilde \e_{m} - \frac{t_{m-1}^2}{  \tilde \Delta_{m-1}(z) }
\end{eqnarray}
where the total coupling matrix element is determined by the integral
\begin{eqnarray}
\label{eqn:vm-integral}
 V^2_m &=& \frac{1}{\pi} \int_{-\infty}^{\infty} d\w\,  {\rm Im} \Delta_{m}(\w-i\delta)
\punkt
\end{eqnarray}
Since the new coupling function $\Delta_{m}(z)$ must be proportional to a Green function, its real part must
vanish for $|\w|\to \infty$ as $1/\w$. Therefore, the energy $\e_{m}$ has to be 
calculated from the first momentum of $\tilde \Delta_{m-1}(z)$
\begin{eqnarray}
\label{eqn:em-integral}
 \tilde \e_m &=& \frac{1}{\pi t_{m-1}^2 } \int_{-\infty}^{\infty} d\w\,  \w  {\rm Im} \tilde \Delta_{m-1}(\w-i\delta)
\komma
\end{eqnarray}
to correctly incorporate the center of mass of the previous reservoir.
Although $\tilde \e_m$  is of the same order as the original NRG Wilson chain parameter $\e_m$
obtained by the standard NRG approach to a non-constant density of states \cite{BullaCostiPruschke2008},
we will show below 
that these values are not identical. In order to be consistent, we need to 
replace $\e_m\to \tilde\e_m$ as given by the first momentum (\ref{eqn:em-integral}). 
Therefore, we will only use the sets of $\{t_m\}$ 
from the NRG approach and replace the Wilson chain
energies accordingly: $\e_m\to \tilde \e_m$.

Let us introduce a 
positive semi-definite but otherwise unspecified cutoff function $F_{d_m}(\w)$ which is
continuous, $ 0\le F_{d_m}(\w)\le 1$, and its smooth transition between 0 and 1
 occurs on the energy scale $d_m$.
For spectral functions $\Gamma_m(\w) = {\rm Im} \Delta_m(w-i0^+)$ with non-zero contributions for positive and negative frequencies, which is the typical situation in the case of fermionic baths 
\footnote{For coupling functions which are non-zero only for $\w>0$ 
as it is the case for bosonic baths \cite{Leggett1987}, $F_{d_m}(\w)$ 
must vanish for all $\w<0$ \cite{SBMopenchain2017}.
}, we demand 
\begin{eqnarray}
F_{d_m}(\w) &\to &
\left\{
\begin{array}{cc}
  1 & \mbox{for}\, |\w|\ll d_m   \\
  0 & \mbox{for}\, |\w|\gg d_m   
\end{array}
\right.
\, .
\end{eqnarray}

We use the cutoff function $F_{d_m}(\w) $ to separate a high-energy part from 
a low-energy part of the coupling function $\Gamma_m(\w)$,
\begin{eqnarray}
\Gamma^L_m(\w) &=& F_{d_m}(\w) \Gamma_m(\w) \non
\Gamma^H_m(\w) &=& \left(1-F_{d_m}(\w)\right)\Gamma_m(\w)
\komma
\end{eqnarray}
so that $\Gamma_m(\w) = \Gamma^L_m(\w) + \Gamma^H_m(\w)$.
This step is schematically shown in Fig.\ \ref{fig:reservoir-iteration-NRG}.
The cutoff energy scale $d_m\propto \lambda^{-m/2}$ must be self-consistently determined by the equation
\begin{eqnarray}
t_m^2  &=& \frac{1}{\pi} \int_{-\infty}^{\infty} d\w \,  \Gamma^L_m(\w-i\delta)
\punkt
\end{eqnarray}
The precise value of $d_m$ will depend on the analytical form of the specific cutoff function $F_d(\w)$.
The separate Hilbert transformation of $\Gamma^L_m(\w) $ and $\Gamma^H_m(\w)$ yields the corresponding real
parts to $\Gamma^L_m(\w) =  {\rm Im} \Delta^L_m(\w-i0^+)$ and $\Gamma^H_m(\w) = {\rm Im} \Delta^H_m(\w-i0^+)$.

Partitioning the new reservoir $m$ into a high and a low energy part,
\begin{eqnarray}
H_{\rm res}(m) &=&H^L_{\rm res}(m)+H^H_{\rm res}(m)
\komma
\end{eqnarray}
the  hybridization to the new chain site $m$ also splits into two parts
\begin{eqnarray}
H_I(m) &=&  H_{I}^L(m) +  H_{I}^H(m) 
\komma
\end{eqnarray}
each involving only low and, respectively, high energy modes:
\begin{eqnarray}
 H_{I}^L(m) &=&
t_m^{ } \left(f^\dagger_m c_{0Lm}^{ } + c_{0Lm}^\dagger f_m^{ } \right)
\\
H_{I}^H(m) &=&
t'_m\left(f^\dagger_m c_{0Hm}^{ } + c_{0Hm}^\dagger f_m^{ }\right)
\punkt
\end{eqnarray}
The high energy coupling constant $t'_m$ accounts for the difference
between $V^2_m$ and $t_m^2$:  $t'_m =\sqrt{V^2_m-t_m^2}$. 
The bath
operators $c_{0Lm}^{ }$ and $c_{0Hm}^{ }$ are a linear combination of these new bath modes
\begin{eqnarray}
c_{0Lm}^{ } &=& \sum_{k}  \lambda_{kLm}^{ } c_{kLm}^{ } 
\non
c_{0Hm}^{ } &=& \sum_{k}  \lambda_{kHm}^{ } c_{kHm}^{ } 
\end{eqnarray}
and also fulfill fermionic commutation relations.
Their corresponding GFs are related to the coupling functions:
\begin{eqnarray}
\correl{c_{0 L m}^{ }}{c^\dagger_{0 L m}}(z) &=& \frac{1}{t_m^2}\Delta^{L}_m(z)
\non
\correl{c_{0 H m}^{ }}{c^\dagger_{0 H m}}(z) &=& \frac{1}{(t'_m)^2}\Delta^{H}_m(z)
\punkt
\end{eqnarray}

After splitting the coupling function $\Delta_m(z)$ into a low and a high energy part,
we use $\tilde\Delta_m(z) = \Delta^L_m(z)$ in the next iteration step $m\to m+1$ 
via Eq.~\eqref{eqn:delta-m-recursion}.  Therefore, 
we have identified the coupling function $\tilde \Delta_{m-1}(z)$ introduced 
in Eq.~(\ref{eqn:define-tilde-delta-m}) as the low energy coupling function of the previous iteration, 
$\tilde \Delta_{m-1}(z)=\Delta^{L}_{m-1}(z)$. 

It should be noted here that $V_m$ is always larger than the desired Wilson chain coupling $t_m$ for any $\Lambda>1$ which ensures that the required reservoirs can be generated for any Wilson chain regardless of the choice of $\Lambda$. $V_0 > t_0 \forall \Lambda > 1$ can be shown analytically (see Appendix \ref{app:reservoirs}). 
If our algorithm generated a $V_m^2<t^2_m$, we would replace $t_m\to V_m$ implying that the chain site $m$ does not couple to an auxiliary high energy  reservoir, i. e.\ $t'_m =0$.

\begin{figure}[t]

 \includegraphics[height=47mm]{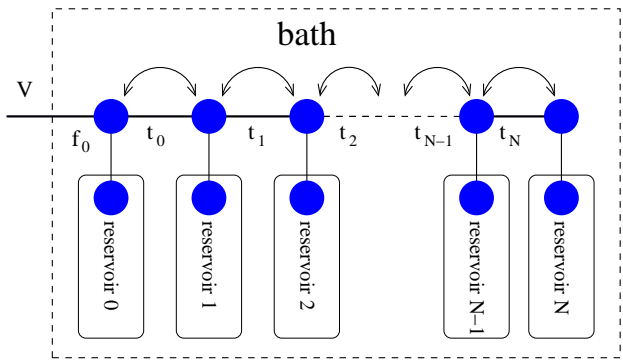}

\caption{The original bath is replaced by a Wilson chain where each chain site $m<N$ 
is coupled to a ``high-energy``
reservoir $H_{\rm res}^H(m)$ and the last site is connected to the remaining reservoir $H_{res}(N)$.}

\label{fig:hybrid-nrg-chain}

\end{figure}

By splitting each coupling function into high and low energy modes the continuous fraction expansion
has been modified such that by coupling a set of additional 
 high-energy reservoirs  $H^{H}_{res}(m)$ 
to the chain site $m$ of a Wilson chain the original continuous coupling function is restored.
The hybrid bath Hamiltonian 
\begin{eqnarray}
\label{eqn:hybrid-nrg-hamiltionian-N-reservoirs}
\tilde H_{\rm bath}(N) &=& H_{\rm chain}(N) +H_{\rm res}(N) + H_{I}(N)
\end{eqnarray}
with the additional reservoirs augmenting the Wilson chain $H_{\rm chain}(N)$
\begin{eqnarray}
H_{\rm res}(N) &=& \sum_{m=0}^{N-1} \sum_{\nu=1}^M H^H_{\rm res,\nu}(m) +
\sum_{\nu=1}^M H_{\rm res,\nu}(N) 
\end{eqnarray}
replaces the original $H_{\rm bath}$ without changing the impurity dynamics. 
This also defines the coupling $H_I(N)$ between the finite size Wilson chain of length $N$ and the
reservoirs
\begin{eqnarray}
\label{eq:H_I_N}
H_I(N) &=&  \sum_{m=0}^{N-1} \sum_{\nu=1}^M H_{I,\nu}^H(m)
 +  \sum_{\nu=1}^M H_{I,\nu}(N).
\end{eqnarray}

Note that we have finally restored the flavor index $\nu$, and the last chain site is coupled
to the full unsplit reservoirs.
The topology of  this resulting hybrid Hamiltonian  is depicted in Fig.~\ref{fig:hybrid-nrg-chain}. 
In the limit  $\Lambda\to 1^+$, $t_{m\nu}^2$ approaches $V_{m\nu}^2$.
As a consequence $t'_{m\nu}\to 0$, and the high energy reservoirs  $H_{I,\nu}^H(m)$  decouple from the system.  In this case, the hybrid Hamiltonian (\ref{eqn:hybrid-nrg-hamiltionian-N-reservoirs})
approaches the DMRG tight-binding chain \cite{KarskiRaasUhrig2008} 
augmented by a single reservoir at the end of the finite size chain.

The  hybrid bath Hamiltonian $\tilde H_{\rm bath}(N)$ consists of the following terms:
the  Wilson chain Hamiltonian $H_{\rm chain}(N)$  generated by the NRG \cite{BullaCostiPruschke2008},
the individual high-energy reservoirs $H^H_{\rm res,\nu}(m)$ at the energy scale $d_m$
and $m<N$, the full remaining reservoir $H_{\rm res,\nu}(N)$ for each flavor $\nu$  and, most
importantly, the coupling between each Wilson chain site $m$ and the corresponding reservoirs $H_I(N)$.\\

\subsection{Reservoir coupling functions $\Gamma_{\nu,m}(\w)$}

\label{sec:reservoir}

In principle, the recursion outlined in the previous section can be applied to any coupling function $\Gamma_\nu(\w)$.
In this paper, however, we restrict ourselves to the simplest case as a starting point of the recursion. Considering a constant density of states within the band $\omega \in [-D,D]$, the hybridization function takes the form 
\begin{equation}
\Gamma(\omega) = \Gamma_0 \Theta(D-|\omega|)	
\label{eq:hybrid_func}
\end{equation}
with the charge fluctuation scale  $\Gamma_0 = \frac{\pi V^2}{2 D}$.
The real part of $\Delta(z)$ is obtained via a Kramers-Kronig relation. Note that we dropped the bath flavor index $\nu$ since we focus on spin degenerate coupling functions in this paper.

\begin{figure}[t]

\includegraphics[width=0.5\textwidth]{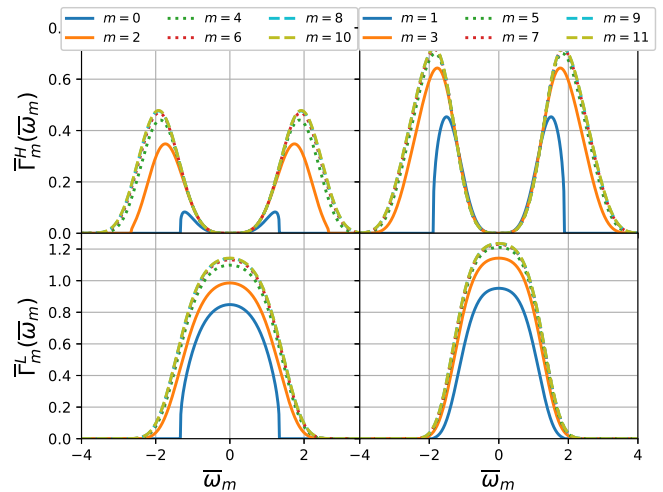}

\caption{Spectral functions $\bar \Gamma_m^{H/L}(\omega) =  {\rm Im} \Delta_m^H(\omega - i0^+)/\w_m$ 
of the high (at the top) and low (at the bottom) energy reservoirs vs $\bar \omega_m =\w/\w_m$.
A bandwidth of $D = 100 \: \Gamma_0$ and a discretization parameter $\Lambda = 2$ have been chosen.}
\label{fig:ReservoirsHigh}
\end{figure}

If at each iteration the reservoir $\Delta(\omega)$ is split into a high energy part $\Delta_H(\omega)$ and a low energy part $\Delta_L(\omega)$ in such a way that the adequate Wilson chain coupling parameters $t_m$ are generated, then the reservoir coupling functions become
invariant at later iterations $m$ if the frequency as well as the magnitude are rescaled by a factor of $\sqrt{\Lambda}$.
The results for these rescaled coupling functions are depicted in Fig. \ref {fig:ReservoirsHigh}. 
The two panels on the l.h.s show the hybridization functions for the even iterations and the two panels on the r.h.s. for the odd iterations respectively.
Clearly, the recursion rapidly approaches convergence.

In deriving the leading order correction to the non-equilibrium dynamics in the presence of these additional reservoirs,
the relaxation matrix acquires contributions of the type $\Gamma_{m}(E_{l_1}^{m_1}-E_{l_2}^{m_2})$, where the coupling function
of the reservoir $m$ must be evaluated at the energy difference between two NRG eigenenergies of two different energy shells $m_1$ and
$m_2$. Taking into account the NRG energy hierarchy, we can conclude from Fig.\ \ref{fig:ReservoirsHigh}
that  $\Gamma_{m}(E_{l_1}^{m_1}-E_{l_2}^{m_2}) \approx 0$, if either 
$m_1< m$ or $m_2< m$.

\section{Non-equilibrium dynamics}
\label{sec:non-equilibrium}

The main focus of this paper is to derive a hybrid approach to the non-equilibirum
dynamics of quantum impurity systems. It combines the time-dependent renormalization
group (TD-NRG) \cite{AndersSchiller2005,AndersSchiller2006} with
a Bloch-Redfield approach \cite{MayKuehn2000} which incorporates the effect of the 
couplings to the additional reservoirs neglected in the NRG onto the real-time dynamics.

\subsection{Non-equilibrium dynamics in the discretized model: the  TD-NRG}

In order to set the stage, we review the TD-NRG which 
is the starting point of the hybrid approach to non-equilibrium. The TD-NRG was derived \cite{AndersSchiller2005,AndersSchiller2006} as an extension of the NRG to access the non-equililibrium dynamics of QIS. The TD-NRG is designed to calculate the full non-equilibrium dynamics of a QIS after a sudden quench: $H(t)=H_0\Theta(-t)+H_f\Theta(t)$ but it is restricted to the discretized representation of the QIS. Recently it was extended to a series  of quenches \cite{CostiFDM-TDNRG2014} mimicking the discretisation of time for a time-dependent  Hamiltonian $H(t)$.

The initial state of the system is assumed to be in thermal equilibrium 
\begin{equation}
{\rho}_0 =
     \frac{ e^{-\beta H_0 }}
          {{\rm Tr}
                \left[
                        e^{-\beta H_0 }
                \right] } .
\label{rho-0}
\end{equation}
At the time $t=0$, the Hamiltonian suddenly switches and the time evolution is governed by  the Hamiltonian $H_f$. We assume that the switching time is short compared to all relevant time scales in the QIS such that it can be viewed as instantaneous. Then, the time evolution of the density operator is given by
\begin{equation}
{\rho}(t > 0) =
    e^{-i t H_f} {\rho}_0
    e^{i t H_f }
\label{rho-of-t}
\end{equation}
for a time-independent $H_f$. Using the complete basis set of the final Hamiltonian the time evolution of any local operator $O$ is given by \cite{AndersSchiller2005,AndersSchiller2006}
\begin{eqnarray}
\langle {O (t)} \rangle &=&
        \sum_{m=m_{\rm min}}^{N}\sum_{r,s}^{\rm trun} \;
        e^{i t (E_{r}^m - E_{s}^m)}
        O_{r,s}^m \rho^{\rm red}_{s,r}(m) ,
\label{eqn:time-evolution-intro} 
\end{eqnarray}
where $E_{r}^m$ and $E_{s}^m$ are the NRG eigenenergies of the Hamiltonian $H_f$ at iteration $m \le N$. $O_{r,s}^m$ is the matrix representation of the operator ${O}$ at that iteration $m$ \cite{BullaCostiPruschke2008}. $m_{\rm min}$ is the first iteration at which the many-body Hilbert space is truncated by the NRG approach. $\rho^{\rm red}_{s,r}(m)$ denotes the reduced density matrix 
\begin{equation}
\rho^{\rm red}_{s,r}(m) = \sum_{e}
          \langle s,e;m|{\rho}_0 |r,e;m \rangle 
\label{eqn:reduced-dm-def}
\end{equation}
in the basis of the final Hamiltonian where the chain degrees of freedom $e$ of the chain sites $m'>m$ (which are called the environment here) are traced out.
In Eq.\ \eqref{eqn:time-evolution-intro} the restricted sums over $r$ and $s$
require that at least one of these states is discarded at iteration $m$: only the discarded states
contribute to the dynamics at iteration $m$.
The kept states $ |k,e;m \rangle$ are refined by adding the chain link couplings  to larger chain sites:
The discarded states at a later iteration are formed from a linear combination
of this tensor product basis. The temperature $T_N \propto \Lambda^{-N/2}$ 
of the TD-NRG calculation is defined by the length of the NRG Wilson chain $N$
and enters Eq. \eqref{rho-0}.

The TD-NRG comprises two simultaneous NRG runs:
one for the initial Hamiltonian $H_0$ in order to compute
the initial density operator ${\rho}_0$ of the system in Eq. \eqref{rho-0} and
one for $H_f$ to obtain the approximate eigenbasis governing the time evolution in Eq. \eqref{eqn:time-evolution-intro}.

This approach has also been extended to multiple quenches \cite{NghiemCosti2014}, time evolution of spectral functions \cite{CostiGF2017} and steady state currents at finite
bias \cite{AndersSSnrg2008,SchmittAnders2009,*SchmittAnders2011,JovchevAnders2013}.
The only error of this method originates from the representation of the bath
continuum by a finite-size Wilson chain \cite{Wilson75} and are
essentially well understood \cite{EidelsteinGuettgeSchillerAnders2012,GuettgeAndersSchiller2013}.

\subsection{Bloch-Redfield extension of the TD-NRG} 
\label{sec:bloch-redfield}

\subsubsection{Introduction}

Finite size oscillations remain present in the TD-NRG expectation value $\langle {O} \rangle (t)$
calculated via Eq.\ \eqref{eqn:time-evolution-intro}
even for $t\to \infty$ depending on the NRG discretization parameters \cite{EidelsteinGuettgeSchillerAnders2012,GuettgeAndersSchiller2013,GuettgePhD2013}. 
We define the averaged steady state value
\begin{eqnarray}
\label{eqn:steady-state-average}
\langle {O} \rangle_\infty &=& \lim_{T\to \infty} \frac{1}{T} \int_0^T dt \langle {O} \rangle (t) 
\non
&=&  \sum_{m=m_{\rm min}}^{N}\sum_{r,s}^{\rm trun} \;
               O_{r,s}^m \rho^{\rm red}_{s,r}(m) \delta_{E_s,E_r}
\end{eqnarray}
predicted by the TD-NRG implying that $r$ and $s$ have to be discarded states. Only the energy diagonal matrix elements contribute to the steady state, which has been extensively discussed in the context of the eigenstate thermalization hypothesis \cite{ETH1991,ETH1994,RigolETH2008,RigolETH2012}. 

Since the contribution of the discarded states of the iterations $m<N$ to the thermodynamic density operator in the NRG is negligibly small,
a thermalized averaged steady state implies vanishing contributions from all $\rho^{\rm red}_{s,r}(m) $ with $m<N$ and an approach of 
$\rho^{\rm red}_{s,r}(N)\to \delta_{E_s^N,E_r^N}\exp(-\beta E_s^N)/Z $.
Within the TD-NRG the values of the matrix elements $\rho^{\rm red}_{s,r}(m)$, however, remain fixed 
and depend on the initial condition \cite{ETH1991,ETH1994,RigolETH2008,RigolETH2012}.
The difference $\Delta O = \langle {O} \rangle_\infty - \langle {O} \rangle_{th}$ quantifies the deviation of the  TD-NRG steady-state prediction from the thermodynamic limit $\langle {O} \rangle_{th}$.

In Sec.\ \ref{sec:chain-plus-single-reservoir} we have proven that the Hamiltonian $\tilde H_{\rm bath}(N)$,
comprising the Wilson chain with $N$ chain links and a sequence of reservoirs, generates the same coupling function
$\Gamma_\nu(\w)$ as the original continuum problem. 
Hence, the Hamiltonian  $H'(N)$,
\begin{eqnarray}
H'(N) &=& H^{\rm NRG}_N   + H_{\rm res}(N) + H_{I}(N)
\end{eqnarray}
is equivalent to the original Hamiltonian $H$ 
\begin{eqnarray}
\label{eq:full-H-N}
H  &=& H_{\rm imp} + H_I +  H_{\rm bath}
\end{eqnarray}
prior to the Wilson discretization 
with respect to its impurity dynamics.
$H'(N)$ augments the standard NRG Hamiltonian of a chain of length $N$, $H^{\rm NRG}_N$, with 
the sum of all additional reservoirs $H_{\rm res}(N)$ and their couplings 
to the chain links $H_{I}(N)$  as stated in detail in Eq.~\eqref{eq:H_I_N}.

The TD-NRG \cite{AndersSchiller2005,AndersSchiller2006} utilizes the standard NRG approximation by replacing 
the original Hamiltonian with the approximation $H\to H^{\rm NRG}_N$. The aim of 
this section \ref{sec:bloch-redfield} is to derive a set of coupled differential equations for the dynamics of
the reduced density matrix $\rho^{\rm red}_{s,r}(m)$ in Eq.\ \eqref{eqn:time-evolution-intro}:
$\rho^{\rm red}_{s,r}(m)\to \rho^{\rm red}_{s,r}(m,t)$. 
The physical origin of the time dependency of the reduced
density matrix is the coupling of the Wilson chain to a set of reservoirs neglected in the NRG approximation.
While the exact solution of $\rho^{\rm red}_{s,r}(m,t)$ in the presence of the additional
reservoirs is complicated and impractical
to implement, we gear towards an approximate solution in the spirit of weak coupling theories
such as a Bloch-Redfield or Lindblad type of master equations \cite{CarmichaelQuantumOpticsI,MayKuehn2000}.

One can explicitly show \cite{CarmichaelQuantumOpticsI,MayKuehn2000}
that  the dynamics of the diagonal elements of the density matrix defined on a finite Hilbert space of dimension $D$
decouples from the off-diagonal dynamics within the  Bloch-Redfield or Lindblad approaches. The
Liovillian operator  
has $D^2$ eigenvalues:  $D$ of them determine the decay into the steady state
while the other $D^2-D$ eigenvalues are complex and always come in pairs $\lambda_i^{ }, \lambda^*_i$, since the density matrix must be hermitian.

Below we derive these two types of differential equations for the diagonal and the off-diagonal
matrix elements of $\rho^{\rm red}_{s,r}(m,t)$. We show that for a generic decay tensor, the diagonal matrix elements approach the thermal equilibrium defined by the full density matrix
formulation \cite{WeichselbaumDelft2007}  of the NRG while the off-diagonal matrix elements vanish in the 
long-time limit.  In order to ensure the conservation of the trace of the density operator, the differential
equation for the diagonal matrix elements requires a coupling of all energy shells, i.e.\ all iterations $m$. 
This sets a practical limit to our approach and additional approximations are required
since the implementation of the couplings between all energy shells is practically impossible.

\subsubsection{Derivation of the second order corrections to the TD-NRG dynamics}
\label{sec:bloch-redfield-operator}

We initially start from the total density operator in the interaction representation  
\begin{eqnarray}
\rho_I(t) &=& e^{iH_0 t} \rho(t)  e^{-iH_0 t}
\end{eqnarray}
where $H_0 = H^{\rm NRG}_N   + H_{\rm res}(N)$. The total density operator
encodes the dynamics of the original problem and operates on the Wilson chain degrees of freedom (DOF) as well
as the DOF of the reservoirs. Neglecting the system-reservoir coupling $H_{I}(N)$ and assuming a factorized density operator in the contributions of
each subsystem yields
a time-independent density operator whose reduced matrix elements relevant for the local expectation values are given by the
TD-NRG values $\rho^{\rm red}_{s,r}(m)$. By incorporating the additional system-reservoir coupling 
the density operator $\rho_I(t)$ acquires the time-dependency
that we cast into a master equation for $\rho^{\rm red}_{s,r}(m,t)$.

The dynamics of the density operator $\rho_I(t)$ 
is governed by the differential equation
\begin{eqnarray}
\partial_t \rho_I(t) &=& i [\rho_I(t), V_I(t)]
\label{VonNeumann}
\end{eqnarray}
in the interaction picture, where the system-reservoir coupling  takes the form
\begin{eqnarray}
V_I(t) &=&e^{ iH_0 t} H_{I}(N) e^{-iH_0 t}.
\end{eqnarray}
For expectation values of local operators it is sufficient to know
$\rho_S(t) = {\rm Tr}_R[\rho_I(t)]$ 
where we have traced out all the reservoir degrees of freedom. This operator is acting only on the Wilson chain
or system $S$.

Now we can adapt Eq.\ \eqref{VonNeumann} to derive
a Bloch-Redfield equation for the reduced density matrix $\rho_S(t)$. 
The individual steps are carried out in appendix \ref{app:bloch-redfield}
and can also be found in textbooks - for example Ref.\ \cite{MayKuehn2000}.

The derivation requires a complete eigenbasis \cite{MayKuehn2000} 
of the discrete system Hamiltonian $H_S$ which is equal to $H_{\rm NRG}(N)$.
For a given NRG eigenbasis $ |r ,e;m \rangle$ of the discrete Hamiltonian $H_S = H^{\rm NRG}_N$,
the Bloch-Redfield master equation reads
\begin{widetext}
\begin{subequations}
\label{MasterEqu}
\begin{align}
\label{MasterEqu-a}
\dot{\rho}_{1,2}(t) &= - \sum_{3,4} R_{1,2;3,4}(t) \rho_{3,4}(t) \\
 R_{1,2;3,4}(t) &= \delta_{2,4} \sum_5 \Xi^+_{1,5,5,3}(t) + \delta_{1,3} \sum_5 \Xi^-_{4,5,5,2}(t) - \Xi^+_{4,2,1,3}(t)  - \Xi^-_{4,2,1,3}(t) 
 \label{MasterEqu-b}
 \\
\Xi^+_{1,2,3,4}(t) &= e^{i(\omega_{1,2}+\omega_{3,4})t} \sum_{\tilde{m}=0}^{N} \sum_\nu \big[C_{\nu,\tilde{m}}(\omega_{3,4})(f^\dag_{\nu,\tilde{m}})_{1,2} (f^{ }_{\nu,\tilde{m}})_{3,4} +\bar C_{\nu,\tilde{m}}(\omega_{3,4})(f^{ }_{\nu,\tilde{m}})_{1,2} (f^{\dag}_{\nu,\tilde{m}})_{3,4} \big] 
\label{MasterEqu-c}
\\
\Xi^-_{1,2,3,4}(t) &= e^{i(\omega_{1,2}+\omega_{3,4})t} \sum_{\tilde{m}=0}^{N} \sum_\nu \big[C_{\nu,\tilde{m}}^*(\omega_{2,1})(f^\dag_{\nu,\tilde{m}})_{1,2} (f^{ }_{\nu,\tilde{m}})_{3,4} + \bar C_{\nu,\tilde{m}}^*(\omega_{2,1})(f^{ }_{\nu,\tilde{m}})_{1,2} (f^{\dag}_{\nu,\tilde{m}})_{3,4} \big], 
\label{MasterEqu-d}
\end{align} 
\end{subequations}
\end{widetext}
with the energy differences $\omega_{i,j}= E_i-E_j$. The index $i \in \{1,2,3,4,5\}$ is a general shortcut notation for the tuple $i=(r_i,e_i,m_i)$, where $ r_i$ is a state label of the NRG state at iteration $m_i$, and $e_i$ is an environment degree of freedom of the remaining $N-m_i$ chain sites. 
$e^{i \omega_{i,j} t} (f^{(\dag) }_{\nu,\tilde{m}})_{i,j}
=\bra{r_i,e_i;m_i} f^{(\dag)}_{\nu,\tilde{m}}(t) \ket{r_i,e_i;m_i}  $ denotes the factorisation of the time-dependent matrix element  of the $\tilde m$-th chain site into a time-independent part and a time-dependent phase factor. The index $\tilde m$ labels the reservoir index of  
the sum over all additional reservoirs in $H_I(N)$.

The bath coupling functions $\Gamma_{\nu m} (\e)$ derived in  Sec.~\ref{sec:chain-plus-single-reservoir} enter the expression as the greater and the lesser GF for each reservoir $G^{>}_{\nu,m}(\tau) / G^{<}_{\nu,m}(\tau)$ and fully determine the effects of the reservoirs onto the dynamics of the Wilson chain. The correlation functions $C_{\nu,m}(\omega)$ and $\bar C_{\nu,m}(\omega)$ are obtained by a half-sided Fourier transformation
\begin{subequations}
\label{eq:reservoir-gfs-FT}
\begin{eqnarray}
C_{\nu,m}(\w) &=& i\int_0^{\infty} d\tau G^{>}_{\nu,m}(\tau) e^{-i\w \tau}\\
\bar C_{\nu,m}(\w) &=& -i\int_0^{\infty} d\tau G^{<}_{\nu,m}(\tau) e^{-i\w \tau}
\end{eqnarray}
\end{subequations}
that results from integrating Eq.\ \eqref{VonNeumann} and then substituting the resulting  expression for $\rho_I(t)$ back into Eq.\ \eqref{VonNeumann}. 
Using  the  definitions
of the lesser and the greater GFs  introduced in Eqs.~\eqref{eq:reservoir-gfs}
we find 
\begin{subequations}
\label{eq:reservoir-Keldysh-relation}
\begin{eqnarray}
C_{\nu,m}(\w) + C^*_{\nu,m}(\w) &=&
 i G^{> }_{\nu,\tilde m}(-\w)
\\
 \bar C_{\nu,m}(\w) + \bar C^*_{\nu,m}(\w) &=& -i G^{> }_{\nu,\tilde m}(\w) 
\end{eqnarray}
\end{subequations}
which relates these combinations to the Fourier transformation of the equilibrium greater and lesser reservoir 
coupling functions.

\subsubsection{Secular approximation}

The objective is to derive a differential equation for the reduced density matrix $\rho^{\rm red}_{s,r}(m;t)$
using Eqs.~\eqref{MasterEqu} and to replace $\rho^{\rm red}_{s,r}(m)$ by its solution 
$\rho^{\rm red}_{s,r}(m;t)$.
The Bloch-Redfield equations introduced in the previous section 
serve as a starting point  for a master equation describing the dynamics of the reduced density matrix
$\rho^{\rm red}_{s,r}(m;t)$ which is defined as 
\begin{eqnarray}
\label{eqn-def-rho-red-m-t}
\rho^{\rm red}_{s,r}(m;t)  &=& \sum_{e}
          \langle s,e;m|{\rho}_S(t) |r,e;m \rangle 
\end{eqnarray}
with $\rho^{\rm red}_{s,r}(m)=\rho^{\rm red}_{s,r}(m;t=0)$ as the initial condition.
The index pair $(s,r)$  can either label  both discarded states or contain only one discarded state, so that
we have to allow for the second state to be retained for the next NRG iteration. Both, however, are approximate
eigenstates of $H_S$, $H_S |r ,e;m \rangle \approx  E_r^m  |r ,e;m \rangle$ and $H_S |s ,e;m \rangle \approx  E_s^m  |s ,e;m \rangle$.

In the next step, we apply  the secular approximation \cite{CarmichaelQuantumOpticsI,MayKuehn2000}.
The remaining explicit time dependency on the  r.h.s of Eq.\ \eqref{MasterEqu-a} in terms of fast oscillating phases, which only occurs in Eqs.\ \eqref{MasterEqu-c} and \eqref{MasterEqu-d}, 
must vanish providing the additional energy constraint 
\begin{eqnarray}
e^{i(\omega_{1,2}+\omega_{3,4})t} \rightarrow \delta_{\omega_{1,2},-\omega_{3,4}}
\label{SecApprox}
\end{eqnarray}
which is consistent with a slowly varying reduced density matrix. 
As a consequence, the time-dependent tensor $R_{1,2,3,4}(t)$ becomes time-independent.

For the dynamics of $\rho^{\rm red}_{s,r}(m;t)$ only the case  $m_1=m_2$  is relevant. The resulting condition $E_{r_1}^{m_1}-E_{r_2}^{m_1} = E_{r_3}^{m_3}-E_{r_4}^{m_4}$ requires the discussion of  two cases (given that degeneracies in $r_1,r_2$ are excluded): 
For the diagonal elements, $r_1=r_2$, immediately  $m_3=m_4$ and $r_3=r_4$ follow 
(since it is highly unlikely to find two different eigenstates at different iterations $m_3\not=m_4$ that are energetically degenerate.)

 If $r_1 \neq r_2$, and thus $E_{r_1}^{m_1}-E_{r_2}^{m_1} \neq 0$, the equation can only be fulfilled for $m_1=m_2=m_3=m_4$, since it is very unlikely to find the same energy difference on two different NRG iterations.

From this discussion we draw two important conclusions:  
(i) For the occupation dynamics given by the diagonal elements of the density matrix $\rho_I^{\rm red}(t)$ (DDM), we obtain Bloch-Redfield tensor matrix elements $R_{1,2,3,4}$ that couple two different iterations $m=m_1=m_2$ and $m'=m_3=m_4$.
(ii) The dynamics of the off-diagonal elements of the density matrix (ODDM) is determined by the coupling
to the reduced density operator within the same energy shell $m$.

\subsection{Dynamics of the reduced density matrix $\rho^{\rm red}_{s,r}(m)$}
\label{sec:neq-reduced-DM}

Within the Bloch-Redfield approach  \cite{MayKuehn2000}  the DDM decouple from the ODDM. 
The DDM describe the occupation dynamics and are coupled by relaxation parameters within the same iteration index $m$ as well as by terms connecting different iterations. These later terms are important for deriving a master equation for the occupation dynamics that satisfies the conservation of the trace of the density matrix at all times.

Guided by the energy separation between the discarded states and the kept states which 
provide the span of the Fockspace for all discarded states at later iterations, we use the approximation
\begin{eqnarray}
\label{eq:approx-rho-red-59}
\bra{r,e;m} \rho_S(t) \ket{s,e';m} \approx \rho^\text{red}_{r,s}(m;t) \delta_{e,e'}d^{-(N-m)}
\label{DensMatApprox}
\end{eqnarray}
for the matrix elements of the reduced density operator $ \rho_S(t)$
which strictly holds only for the equilibrium density operator \cite{WeichselbaumDelft2007}. 
Once we trace out the environment DOF $e$, the factor $d^{-(N-m)}$ is canceled and the definition of $\rho^{\rm red}_{r,s}(m;t) $ introduced in Eq.~\eqref{eqn-def-rho-red-m-t} is recovered.\\

\subsubsection{Diagonal part of the reduced density matrix}

\label{sec:occupation-dynamics}

To evaluate the DDM, $1=2, 3=4$ 
has to be set in Eq.\ \eqref{MasterEqu} to arrive at
\begin{eqnarray}
\dot{\rho}^\text{red}_{l_1,l_1}(m_1;t) &=& \sum_{l_2,m_2} \Big(\Xi_{l_2,l_1}(m_2,m_1) \rho^\text{red}_{l_2,l_2}(m_2;t) 
\label{DiagPart}
\\
&& \phantom{\sum_{l_2,m_2}}  - \Xi_{l_1,l_2}(m_1,m_2) \rho^\text{red}_{l_1,l_1}(m_1;t) \Big)
\nonumber
\end{eqnarray}
with the relaxation matrix elements
\begin{eqnarray}
\Xi_{l_1,l_2}(m_1,m_2) &=& d^{-(N-m_1)}\sum_{e_1,e_2} \left(\Xi_{1,2,2,1}^+ + \Xi_{1,2,2,1}^-  \right).
\nonumber
\end{eqnarray}

Eq.~\eqref{DensMatApprox} demands that $e_1=e_2$ as well as $e_3=e_4$ in Eq.\ \eqref{MasterEqu}. 
Thus only terms of the form $\bra{r,e;m} \rho_S(t) \ket{s,e;m}$ occur in Eq.\ \eqref{MasterEqu}, and  the environment $e_1$ has been traced out on both sides of Eq.\ \eqref{DiagPart}.
Note that the DDM are restricted to the discarded states $l_i$ of the iteration $m_i$, since 
the complete basis set used to evaluate the trace of the density matix comprises all discarded states
\cite{AndersSchiller2005,AndersSchiller2006} and a combination of two kept states does not contribute in Eq.~\eqref{eqn:time-evolution-intro}.

For the DDM, the relations between the different  half-sided Fourier components in systems with identical chemical potentials in each reservoir
\begin{align*}
C_{\nu,\tilde{m}}(\omega_{2,1}) + C_{\nu,\tilde{m}}^*(\omega_{2,1}) &= 2 f(\omega_{2,1}) \Gamma_{\nu,\tilde{m}}(\omega_{1,2})\\
\overline{C}_{\nu,\tilde{m}}(\omega_{2,1}) + \overline{C}_{\nu,\tilde{m}}^*(\omega_{2,1}) &= 2 f(\omega_{2,1}) \Gamma_{\nu,\tilde{m}}(\omega_{2,1})
,
\end{align*}
are used -- see also Eq.\ \eqref{eq:reservoir-Keldysh-relation} --
to derive the explicit expression of the relaxation tensor matrix elements
\begin{widetext}
\begin{align}
\label{eq:Xi_m1_m2}
\Xi_{l_1,l_2}(m_1,m_2) &= 
\frac{2 f(\omega_{2,1})}{d^{N-m_1}} 
 \left(W_{l_1,l_2}^{(m_1,m_2)} + W_{l_2,l_1}^{(m_2,m_1)} \right) \\
 \label{def-w-l1-l2}
W_{l_1,l_2}^{(m_1,m_2)}&= \sum_{\tilde{m} = 0}^{M} \sum_\nu  \Gamma_{\nu,\tilde{m}}(\omega_{1,2})X^{\tilde m}_{l_1,l_2}(m_1,m_2)\\
\label{XiDiag}
X^{\tilde m}_{l_1,l_2}(m_1,m_2) &=  \sum_{e_1,e_2} \bra{l_1,e_1;m_1} f^\dag_{\nu,\tilde{m}} \ket{l_2,e_2;m_2} \bra{l_2,e_2;m_2} f^{ }_{\nu,\tilde{m}} \ket{l_1,e_1;m_1} ,
\end{align}
\end{widetext}
where in general the number of reservoirs is determined by the chain length, i.\ e.~$M=N$. 
The first term on the r.h.s of \eqref{eq:Xi_m1_m2} describes  the emission of a particle into
the reservoir $\tilde m$ and afterwards a reabsorbition
while the second term starts with an absorption and ends with a reemission process.

It is easy to check that the sum $W_{l_1,l_2}^{(m_1,m_2)} +  W_{l_2,l_1}^{(m_2,m_1)}$
is symmetric with respect to exchanging the label pairs $(l_1,m_1)\leftrightarrow (l_2,m_2)$.
Therefore, the asymmetry in the rates $\Xi_{l_1,l_2}(m_1,m_2)$  with respect to this index 
swap is solely caused by the prefactor.

The steady-state value of the reduced density matrix is fully determined by the prefactor 
$f(\omega_{2,1}) d^{m_1-N}$.
The specific form of the remaining term $W_{l_1,l_2}^{(m_1,m_2)} +  W_{l_2,l_1}^{(m_2,m_1)}$ is irrelevant for the steady-state values and only influences the relaxation time scales as long as all matrix elements remain coupled in this master equation. Therefore, a decoupling  of bound states on the Wilson chain from the reservoir continuum would lead to a steady-state of the system which deviates from the thermal equilibrium.

We discuss two important properties of the master equation \eqref{DiagPart}.
Firstly, the trace $\text{Tr}[\rho_S] = \sum_{l,m} \rho^\text{red}_{l,l}(m;t)$ is conserved at all times $t$, since 
\begin{eqnarray}
0 &=& \partial_t \text{Tr}[\rho_S] = \sum_{l_1,m_1} \dot{\rho}^\text{red}_{l_1,l_1}(m_1;t) \\
&=& \sum_{l_1,m_1} \sum_{l_2,m_2} 
\Big(  \Xi_{l_2,l_1}(m_2,m_1) \rho^\text{red}_{l_2,l_2}(m_2;t) 
\non
&& \phantom{\sum_{l_1,m_1} \sum_{l_2,m_2} }
- \Xi_{l_1,l_2}(m_1,m_2) \rho^\text{red}_{l_1,l_1}(m_1;t) \Big).
\nonumber
\end{eqnarray}
This can be seen by interchanging the summation indices $(l_1,m_1)$ and $(l_2,m_2)$  in the second summation.

Secondly, the steady state of the matrix elements obeys the detailed balance condition.
Since $f(\omega_{2,1}) e^{-\beta E_{l_1}} = f(\omega_{1,2}) e^{-\beta E_{l_2}}$ holds, and thus 
$ \Xi_{l_1,l_2}(m_1,m_2) e^{-\beta E_{l_1}} 
= \Xi_{l_2,l_1}(m_2,m_1) e^{-\beta E_{l_2}} 
d^{m_1-m_2}$, the fixed point of Eq.~\eqref{DiagPart} 
is given by 
\begin{eqnarray}
\label{eq:rho-fixed-point}
\rho^{\text{red}}_{l,l}(t\rightarrow \infty;m) &= &\frac{d^{N-m}}{Z} e^{-\beta E_{l}^m}
\end{eqnarray}
with the partition sum $Z$ \cite{WeichselbaumDelft2007} 
\begin{eqnarray}
Z &= & \sum_{m=m_\text{min}}^N \sum_l d^{N-m} e^{-\beta E_l^m}.
\end{eqnarray}
The formalism requires that $E_l^m$ is given in the absolute energy units measured relative to the ground state energy on the last iteration $E_g^{N}$, which  comprises the sum of the rescaled NRG  eigenenergies $\bar E_l^m$ and the ground state energy shift relative to the last iteration, $\Delta E_g^{N,m} = E_g^{m}-E_g^{N}$. Since the ground state energy is  reduced in each iteration step, a positive constant is added to $E_l^{m} = \Lambda^{(m-1)/2}\bar E_l^m$ which in combination with the low temperature $1/\beta$ causes an exponential suppression of the contributions for $m<N$ even for $\bar E_l^m= 0$ on the specific iteration $m$ after identifying
$\beta=\beta_N \propto \Lambda^{(N-1)/2} \bar \beta$ \cite{BullaCostiPruschke2008}.

The steady-state  fixed point stated in Eq.\ \eqref{eq:rho-fixed-point} is independent of the values of $X_{l_1,l_2}(m_1,m_2) $ unless some  matrix elements vanish. Therefore, $\rho^\text{red}_{l,l}(m;t)$  in general approaches its thermal equilibrium value. 
If, however, the reservoirs have different chemical potentials
this statement does not hold. In that case the structure of the master equation suggests the approach 
to a steady state that differs from thermal equilibrium \cite{NussArrigoni2015,DoraArrigioni2015}.

The calculation of all matrix elements $\Xi_{l_1,l_2}(m_1,m_2)$ for all combinations of discarded states between all iterations $m_1,m_2$ is numerically very expensive and appears to be not feasible. Therefore, we hereinafter propose further approximations that 
do not violate the conservation of the trace as well as the thermalization of the density matrix but
keep the approach manageable even for large Fockspaces.

Firstly, we restrict the summation of the reservoirs in 
Eq.\ \eqref{MasterEqu-c}, Eq.\ \eqref{MasterEqu-d} and in particular in Eq.\ \eqref{eq:Xi_m1_m2} 
to $\tilde m \le M=\text{min}(m_1,m_2)$. 
This is a consequence of the analytic properties of the coupling functions $\Gamma_{\nu,\tilde{m}}(\omega)$
discussed at the end of Sec.~\ref{sec:reservoir}.

\subsubsection{Calculation of the matrix elements $\Xi_{l_1,l_2}(m_1,m_2)$}
\label{sec:redfield-diagonal}

The key ingredient of the master equation is the calculation of the transition rates
$\Xi_{l_1,l_2}(m_1,m_2)$ as defined in Eq.\ \eqref{eq:Xi_m1_m2}. While it is straight forward
to evaluate the expressions for $m_1=m_2$, it is a challenge to connect different Wilson shells.
Therefore, we focus on $m_1\not = m_2$ in the following.

We make use of the NRG hierarchy implying that
$f(E_{l_2}^{m_2}- E_{l_1}^{m_1})\approx \Theta(m_2-m_1)$. This implies that the density matrix element
$\rho^\text{red}_{l_1,l_1}(m_1;t)$ in Eq.\ \eqref{DiagPart} decays only into states 
with smaller energies, i.\ e.\ $m_2 \ge m_1$. The first term on the r.h.s of this equation is a source term which increases
the occupation of the state $l_1$ via the decay of states $l_2$ from iterations $m_2\le m_1$.

Using the properties of the coupling functions $\Gamma_{\nu,\tilde{m}}(\w)$ further justifies the simplification 
\begin{eqnarray}
\Gamma_{\nu,\tilde{m}}(\pm \Delta E)
&\approx &
\left\{
\begin{array}{ccc}
 \Gamma_{\nu,\tilde m}(\mp E_{l_1}^{m_1})  &    \mbox{for} \, \tilde m\le m_1
 \\
 0 &    \mbox{for} \, \tilde m> m_1.
\end{array}
\right.
\end{eqnarray}
In order to proceed, we use 
\begin{equation}
1_m^- = \sum_{m' = m_{\rm min}}^{m}
               \sum_{l,e}
                    |l ,e;m'\rangle\
                    \langle l,e;m'|
\label{eqn:cal-I_m^-}
\end{equation}
and
\begin{equation}
1_m^+ = \sum_{k,e}
               |k,e;m\rangle \langle k,e;m| \; .
\label{eqn:cal-I_m^+}
\end{equation}
to partition the completeness relation \cite{AndersSchiller2005,AndersSchiller2006}
\begin{equation}
1 = {1}_m^{-} + {1}_m^{+} \; ,
\label{eqn:completenes-iteration-m}
\end{equation}
of the Fockspace of the Wilson chain. Since discarded states at a later iteration $m_2>m_1$
only have an overlap with the kept states after the iteration $m_1$, we need to evaluate
\begin{eqnarray}
X^{\tilde m}_{l_1,l_2}(m_1,m_2) &=&  \sum_{e_1,e_2} 
\sum_{k,e}\sum_{k',e'}
\bra{l_1,e_1;m_1} f^\dag_{\nu,\tilde m} 
               |k,e;m_1\rangle 
\non
&&
\times \langle k,e;m_1| l_2,e_2;m_2\rangle 
\bra{l_2,e_2;m_2} k',e';m_1\rangle 
\non
&&
\langle k',e';m_1| f^{ }_{\nu,\tilde m} \ket{l_1,e_1;m_1} ,
\end{eqnarray}
from which $X^{\tilde m}_{l_2,l_1}(m_2,m_1)$ can be derived by exchanging the 
operators $f$ and $f^\dag$.

Then the matrix elements of the creation and annihilation  operator
are diagonal in the environment variables $e_1,e$ and $e'$: 
\begin{eqnarray}
\bra{l_1,e_1;m_1} f^\dag_{\nu,\tilde m} 
               |k,e;m_1\rangle &=& \delta_{e_1,e} ( f^\dag_{\nu,\tilde m})_{l_1,k}
               \\
   \langle k',e';m_1| f^{ }_{\nu,\tilde m} \ket{l_1,e_1;m_1}  &=&        
   \delta_{e_1,e'} ( f_{\nu,\tilde m})_{k',l_1}     
\end{eqnarray}
leaving the calculation of the general overlap matrix elements
\begin{eqnarray}
\label{eq:64-A}
S_{l_2^{ },l'_2;k,k'}^{(m_1,m_2)}
 &=&
\sum_{e_1,e_2}  \langle k,e_1;m_1| l_2,e_2;m_2\rangle 
\non
&& \times
\bra{l'_2,e_2;m_2} k',e_1;m_1\rangle 
\end{eqnarray}
where we set $l_2=l_2'$ at the end. This can most easily be evaluated in terms of a matrix product
formulation \cite{Schollwoeck2011}.

\begin{figure}[t]
\begin{center}
\includegraphics[width=0.48\textwidth]{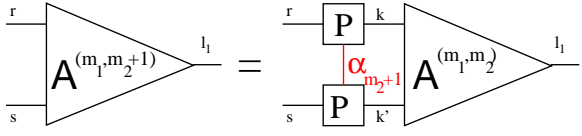}

\caption{Diagrammatic representation of the recursion
    relation for calculating the $\mat{A}$ tensor.
     Each box represents a matrix element $P_{r_{m_2+1},k_{m_2}}[\alpha_{m+1}]$ 
   (upper
    row) or its complex conjugate (lower row). The state
    labels $k$ and $k'$ are plotted horizontally. The state
    label $\alpha_{m_2+1}$ for the $m+1$ site is plotted
    vertically. A connected line indicates a summation over
    the corresponding index.
    In analogy to Fig.\ 2 in Ref.\ \cite{AndersSchiller2006}.  
    }
\label{fig:A-recusion}
\end{center}
\end{figure}

We recall that the NRG eigenstates at the iteration $m_1+1$ can be expanded as
\begin{eqnarray}
\ket{r,e;m+1} &=& \sum_{k,\alpha} P^{m}_{r,k}[\alpha] \ket{k,\alpha,e;m}
\end{eqnarray}
where $k$ denotes the kept states after the iteration $m$ and $\alpha$ labels the DOF of the
chain site $m+1$. The matrix $P^{m}_{r,k}[\alpha]$ is generated during the diagonalization of 
$H_{m+1}^{\rm NRG}$. Recursively applying this relation leads to the matrix product expansion
\begin{eqnarray}
| l_2,e_2;m_2\rangle 
&=&  \sum_{k} \sum_{\{\alpha_i \} } \prod_{i=m_1}^{m_2-1}
 \mat{P^i}[\alpha_{i}]  \ket{k,\{\alpha_i\} ,e_2;m_1 }
\end{eqnarray}
which we insert into \eqref{eq:64-A} to obtain the overlap tensor
\begin{eqnarray}
\label{eq:S-71}
S_{l_2^{ },l'_2;k,k' }^{(m_1,m_2)} &=&
d^{N-m_2}
\sum_{\{\alpha_i \} }
\left[
\prod_{i=m_1}^{m_2-1} \mat{P^i}[\alpha_{i}]
 \right]_{l_2,k}
\non
 && \times
\left[
\prod_{i=m_1}^{m_2-1} \mat{P^i}[\alpha_{i}]
 \right]^*_{l'_2,k'}.
 \end{eqnarray}
The prefactor $d^{N-m_2}$ arises from performing the summation over the remaining diagonal environment DOF.

The calculation of $S_{l_2,l'_2;k,k' }^{(m_1,m_2)}$
can be casted  in the recursion relation
\begin{eqnarray}
\label{eq:S-recusion}
S_{ l_2^{ },l'_2;k,k'}^{(m_1,m_2+1) }
&=& \frac{1}{d} \sum_{\alpha_{m_2+1}}
\sum_{k_1,k_2} P^{m_2+1}_{l_2,k_1}[ \alpha_{m_2+1}] 
\non
&& \times
\left[ P^{m_2+1}_{l_2,k_1}[ \alpha_{m_2+1}]\right]^*
 S_{k_1,k_2; k,k' }^{(m_1,m_2)} \punkt
\end{eqnarray}
Although this expression can be diagrammatically visualised in terms of matrix product states \cite{Schollwoeck2011}.
such a tensor with six indices is numerically not manageable and can only serve as an auxiliary quantity. 

The recursion relation of the tensor $S_{ l_2^{ },l'_2;k,k'}^{(m_1,m_2+1) }$, however, allows to derive a
recursion relation for the decay rates  $W_{l_1,l_2}(m_1,m_2)$ defined in Eq.\ \eqref{def-w-l1-l2}.
For that purpose we introduce the tensor
\begin{eqnarray}
\label{eq:Def-Fkk'l}
F_{k,k'}(l_1,m_1) &=& \sum_{\tilde{m} = 0}^{m_1} \sum_\nu  \Gamma_{\nu,\tilde{m}}(E_{l_1}^{m_1})
( f^\dag_{\nu,\tilde m})_{l_1,k} \\
&& \phantom{\sum_{\tilde{m} = 0}^{m_1} \sum_\nu } 
\times ( f_{\nu,\tilde m})_{k',l_1}    .
\nonumber
\end{eqnarray}
This includes all reservoir coupling functions up to $\tilde m\le m_1$. Due to the analytic properties of $ \Gamma_{\nu,\tilde{m}}(\w)$
we expect that $\Gamma_{\nu,\tilde{m}}(E_{l_1}^{m_1})$ is rapidly vanishing for $\tilde m \ll
m_1$ so that $\tilde m=m_1$ will be the major contribution. From the definition of $W_{l_1,l_2}(m_1,m_2) $ we immediately 
obtain
\begin{eqnarray}
W_{l_1,l_2}(m_1,m_1) &=& d^{N-m_1}  F_{l_2,l_2}(l_1,m_1) 
\end{eqnarray}
for the Bloch-Redfield tensor elements connecting states
on the same Wilson shell $m_1=m_2$. The prefactor $d^{N-m_1}$ arises from the trace over the remaining environment DOFs
and compensates the prefactor $d^{-(N-m_1)}$ in $\Xi_{l_1,l_2}(m_1,m_2)$.  
Let us absorb the prefactor $d^{-(N-m_1)}$ in the definition 
\begin{eqnarray}
\frac{1}{d^{N-m_1}} W_{l_1,l_2}(m_1,m_2) = A_{l_2,l_2;m_2}(l_1,m_1)
\end{eqnarray}
where the tensor $A_{r,s;m_2}(l_1,m_1)$ is given by the contraction 
of the overlap tensor $\mat{S}$ and the coupling tensor $\mat{F}(m_1)$
\begin{eqnarray}
A_{r,s;m_2}(l_1,m_1) &=& \frac{1}{d^{N-m_1}} \sum_{k,k'}  
 S_{r,s; k,k' }^{(m_1,m_2)}\\
 && \times  F_{k,k'}(l_1,m_1)
\nonumber
\end{eqnarray}
This $\mat{A}$-tensor obeys the recursion
\begin{eqnarray}
\label{eq:A-recusion}
A_{r,s;m_2+1}(l_1,m_1)
&=& \frac{1}{d} \sum_{\alpha_{m_2+1}}
\sum_{k_1,k_2} P^{m_2+1}_{l_2,k_1}[ \alpha_{m_2+1}] 
\\
&& \times
\left[ P^{m_2+1}_{l_2,k_1}[ \alpha_{m_2+1}]\right]^*
A_{k_,k_2;m_2}(l_1,m_1)
\nonumber
\end{eqnarray}
using the tensor $F_{k,k'}(l_1,m_1)$ as the initial condition,
derived from the recursion \eqref{eq:S-recusion}. This recursion is visualized in  Fig. \ref{fig:A-recusion}.

Since the coupling tensor $\mat{F}(m_1)$ has been included in the definition, the $A$ tensor has three indices for each combination  $(m_1,m_2)$ of iterations.
Note that the prefactor in $S(m_1,m_2)$, $d^{N-m_2}$, can be combined with
the overall prefactor of $\Xi_{l_1,l_2}(m_1,m_2)$, $d^{-(N-m_1)}$, to obtain 
$d^{-(m_2-m_1)}$ which only depends on the relative distance between the iterations. 
After calculating $A_{r,s;m_2}(l_1,m_1)$ for all states $r,s$ present at iteration $m_2$,
the diagonal matrix elements of discarded states, $A_{l_2,l_2;m_2}(l_1,m_1)$, 
enter the master equation while the kept sector, $A_{k,k';m_2}(l_1,m_1)$, is used in the recursion \eqref{eq:A-recusion}.

Inspecting of $ X^{\tilde m}_{l_2,l_1}(m_2,m_1)$
in the definition \eqref{XiDiag} reveals that the only difference in the calculation is the combination of
annihilation and creation operators. We include this difference into
the tensor $\tilde F$ 
\begin{eqnarray}
\tilde F_{k,k'}(l_1,m_1) &=& \sum_{\tilde{m} = 0}^{m_1} \sum_\nu  \Gamma_{\nu,\tilde{m}}(-E_{l_1}^{m_1})
( f_{\nu,\tilde m})_{l_1,k} \\
&& \phantom{\sum_{\tilde{m} = 0}^{m_1} \sum_\nu } 
\times ( f^\dag_{\nu,\tilde m})_{k',l_1}    
\nonumber
\end{eqnarray}
which differs from Eq.\ \eqref{eq:Def-Fkk'l} by the exchange of matrices for
$f\leftrightarrow f^\dagger$ and the sign of the energy.
By adding $\tilde F_{k,k'}(l_1,m_1)$ and $F_{k,k'}(l_1,m_1)$
and using this sum as initial condition in Eq. \eqref{eq:A-recusion}
generates recursively the sum $W_{l_1,l_2}^{(m_1,m_2)} +\tilde W_{l_1,l_2}^{(m_1,m_2)}$
after setting $r,s=l_2$.

\subsubsection{Approximations of the rates for the diagonal master equation}
\label{sssec:approx_of_rates}

Although the calculation of each matrix element for the diagonal parts of the Bloch-Redfield tensor is analytically straight forward and can be casted into the diagrammatical  matrix product state recursion depicted in Fig.\ \ref{fig:A-recusion},
we want to point out that one needs a third-order tensor $A_{r,s;m_2}(l_1,m_1)$ at any time of the calculations.  
Although the recursions for calculating the sequence of tensors  $A_{r,s;m_2}(l_1,m_1)$
for a fixed value $m_1$ can be independently evaluated for each start iteration $m_1$, running these calculations in parallel requires a large number of such tensors in the memory at any given time.
Therefore, it might be more feasible to run the recursion for each $m_1$ sequentially and use highly parallelized matrix multiplication libraries.

However, in this paper we have chosen a different approach.
Consider that the correct Boltzmann distribution is enforced by the prefactor of $\Xi_{l_1,l_2}(m_1,m_2)$, 
$f(E_{l_2}^{m_2} - E_{l_1}^{m_1})/d^{N-m_1}$, which ensures that the thermodynamic state is always reached. The factor $2(W_{l_1,l_2}^{(m_1,m_2)} + W_{l_2,l_1}^{(m_2,m_1)})$ only determines the relaxation time scale.
We recall that the deviation of the TD-NRG steady state and the NRG thermodynamic expectation value is usually small and within  1-10\%. Therefore, the main purpose of the master equation \eqref{DiagPart} is to ensure the decay of the diagonal matrix elements into the thermodynamic steady state while maintaining the correct decay rate.  
Since the Redfield tensor  decays exponentially with  increasing distance $|m_1-m_2|$, we calculate $\Xi_{l_1,l_2}(m_1,m_2)$ exactly only for the tridiagonal terms $(m_1,m_2 \in \{m_1-1, m_1,  m_1+1 \})$. For $|m_1-m_2|>1$, we replace the exact value of $X_{l_1,l_2}^{\tilde m}(m_1,m_2)$ in Eq. \eqref{XiDiag} by $d^{N- \rm max{(m_1,m_2)}} \delta_{Q_1, Q_2+1}$ where $Q_i$ is the particle number of the state $\ket{l_i,m_i}$. This approximation includes the degeneration of states with the environment parameter $e_i$ as well as the fact that only those states couple whose numbers of particles on the Wilson chain differ by one. In other words: we ignore the correct overlap matrix elements but include the proper symmetry relation between $l_1$ and $l_2$ which demands that transitions are only allowed if the states can be linked by an absorption or an emission of a particle from or into the reservoir.

\subsubsection{Off-diagonal part of the density matrix}

As a consequence of the secular approximation in Eq. \eqref{SecApprox}, only the states of the same NRG iterations $m$ are coupled for the ODDM.
As explained above, it is highly unlikely that the same finite energy difference of the two states $r,s$ at iteration $m$
can be found at any other iteration $m'$ given the energy hierarchy of the NRG approach.
Then Eq.~\eqref{MasterEqu} simplifies to
\begin{align}
\dot{\rho}^\text{red}_{r_1,r_2}(m;t) = - \sum_{r_3,r_4} 
R_{r_1,r_2; r_3,r_4} (m)
\rho_{r_3,r_4}^\text{red}(m;t)
\label{ODDMMasterEqu}
\end{align}
where the environment variables $e_i$ have been traced out canceling the 
factor  $d^{-(N-m)}$ in Eq.\ \eqref{eq:approx-rho-red-59}.

The ODDM has to vanish in the limit $t\rightarrow \infty$ to allow for the correct thermalization. 
This condition is met by the solution of Eq.~\eqref{MasterEqu}. By definition $1 \neq 2$ and $3\neq 4$  must hold: The only possible fixed point of Eq.~\eqref{ODDMMasterEqu} is $\rho^{\text{red}}_{r_1,r_2}(t\rightarrow \infty;m) = 0$ for all $r_1 \neq r_2$.

The calculation of the Bloch-Redfield tensor $R_{r_1,r_2;r_3,r_4}$ defined in Eq.\ \eqref{MasterEqu-b} involves intermediate states which run over the complete basis set of the Wilson chain. Using the Eqs.\ (\ref{eqn:cal-I_m^-}-\ref{eqn:completenes-iteration-m}) allows to divide the intermediate sum over the index $5$ in the two first terms in Eq.\ \eqref{MasterEqu-b} into contributions from the same Wilson shell and contributions from $m'<m$ generated by $1_m^-$. Neglecting the latter contributions retains the structure of the master equation for the ODDM and only leads to a slight underestimation of the relaxation rates \footnote{Note that the fixed point  $\rho^{\text{red}}_{r_1,r_2}(t\rightarrow \infty;m) = 0$ for all $r_1 \neq r_2$ remains unaltered.}. In favor of a fast and simple implementation we only include matrix elements of $\Xi^{\pm}_{1,2,3,4}$ where all four indices are referring to states at the same shell and used the definitions \eqref{MasterEqu-c}
and  \eqref{MasterEqu-d}.

\subsubsection{Combined approach}
\label{sssec:Hybrid}

In the previous sections, we derived the master equation for the reduced
density matrices  $\rho^{\rm red}_{s,r}(m,t)$ that will replace the time-independent reduced
density matrices in Eq.\ \eqref{eqn:time-evolution-intro} by our proposed hybrid TD-NRG approach
\begin{eqnarray}
\langle {O(t) } \rangle &=&
        \sum_{m=m_{\rm min}}^{N}\sum_{r,s}^{\rm trun} \;
        e^{i t (E_{r}^m - E_{s}^m)}
        O_{r,s}^m \rho^{\rm red}_{s,r}(m,t) 
\label{eqn:hybrid-time-evolution-intro} 
\end{eqnarray}
which is the main result of this paper. 

For the conservation of the trace, all reduced density matrix elements of the discarded states
need to be coupled and it is crucial to maintain the symmetry of the Redfield tensor matrix elements $\Xi_{l_1,l_2}(m_1,m_2)$
in Eq.\ \eqref{DiagPart}. 
At any time, the condition
\begin{eqnarray}
\sum_{m=m_{\rm min}}^{N} \sum_l \rho^{\rm red}_{l,l}(m,t) &=& 1
\end{eqnarray}
must hold, where $l$ only includes the discarded states at iteration m.
We fulfill this requirement by solving a master equation for the diagonal matrix elements of the reduced density matrix, Eq.\ \eqref{DiagPart}, as a first step. 
The off-diagonal dynamics only involves couplings within a single Wilson shell and is obtained in a second step.
In a third step, the solutions for $\rho^{\rm red}_{s,r}(m,t)$ are inserted into Eq.\ \eqref{eqn:hybrid-time-evolution-intro}, and the non-equilibrium dynamics of the quantity of interest is evaluated.

\subsection{Algorithms for solving the master equations}
\label{sec:algorithm}

The master equations Eq.\ \eqref{DiagPart} and Eq.\ \eqref{ODDMMasterEqu} are transformed into a Lindblad-style master equation that can be solved by diagonalizing the occurring nonsymmetric matrix. For a long NRG chain with a large number $N_s$ of retained NRG eigenstates, the exact diagonalization of this nonsymmetric matrix is not possible, and we have to rely on approximate schemes. For that purpose the biorthogonal Lanczos algorithm is utilized.

\subsubsection{The Lindblad master equation}

The DDM and the ODDM yield two separate equations that are solved separately. 
In both cases, the reduced density matrices $\rho_{r,s}^\text{red}(m;t)$ are transformed into a super vector that contains all matrix elements. We map the diagonal matrix elements and off-diagonal density matrix elements onto equivalent vectors $ \rho^\text{red}(t) \to \vec{\rho}_{\rm DDM}(t), \vec{\rho}_{\rm ODDM}(t)$  \cite{Dzhioev_2012} and identify the corresponding relaxation matrix. For both cases, we cast the master equations into the form
\begin{eqnarray}
\dot{\vec{\rho}}(t) = - R \: \vec{\rho}(t) \punkt
\label{eq:BRT}
\end{eqnarray}

For the DDM all NRG iterations are connected, whereas in the case of the ODDM only the matrix elements of the  same shells couple to each other. However, the dimension of the master equation of the ODDM comprises two NRG state indices $r,s$ of the same iteration so that the dimension of the off-diagonal vector $\vec{\rho}_{\rm ODDM}$ is $d^2 N_s^2$, where $N_s$ denotes the number of kept states a after each iteration and $d$ the number of local degrees of freedom added in the next iteration step.

$R$ is always a nonsymmetric matrix, and thus we have to distinguish left eigenvectors $\vec{w}_k$ and right eigenvectors $\vec{v}_k$ \cite{SaadSparseLinearSystems2003} 
\begin{eqnarray}
R\vec{v}_k &=& \lambda_k \vec{v}_k\\
\vec{w}^h_k R\ &=& \vec{w}^h \lambda_k
\end{eqnarray}
where $\lambda_k$ is an eigenvalue of $R$.
It should be stressed here, that the eigenvectors $\{ \vec{w}_k, \vec{v}_k \}$ constitute a biorthogonal basis, which is a consequence of the fact that the matrix $R$ is nonsymmetric. 
The eigenvectors obey the biorthogonality relation $\langle \vec{w}_k, \vec{v}_{k'} \rangle = \vec{w}^h_k \vec{v}_{k'} =\delta_{k,k'}$.
Note that right eigenvectors are not orthogonal to each other, and $\langle \vec{w}, \vec{v} \rangle$ denotes the abstract scalar product.

The master equations can be formally solved by
\begin{eqnarray}
\vec{\rho}(t) = e^{- R t} \vec{\rho}(t=0) = \sum_{k=1}^{D} c_k e^{- \lambda_k t} \vec{v}_k \komma
\label{eq:LindbladED}
\end{eqnarray}
where $D$ is the dimension of the density matrix vector $\vec{\rho}(t)$, and the complex expansion 
coefficients $c_k$ are calculated by the scalar product $c_k = \langle \vec{w}_k, \vec{\rho}(t=0) \rangle$.
The supervector $\vec{\rho}(t)$ consists either of the diagonal matrix $\rho^\text{red}_{l,l}(m;t)$ spanning all iterations $m \in[m_{\rm min}, N]$ or, the off-diagonal matrix $\rho^\text{red}_{r,s}(m;t) (r\not= s)$ for each iteration $m$ and is provided by the TD-NRG algorithm. The sum over $k$ comprises a full basis of eigenvectors of $R$, thus Eq.\ \eqref{eq:LindbladED} is exact.

\subsubsection{The biorthogonal Lanczos method}

Since the matrix dimension of the Redfield tensors scale as $N_s^4$ and are much too large for exact diagonalization in a typical NRG framework, we have to employ a Lanczos algorithm to obtain approximate eigenvalues and -vectors in a space of reduced dimension. The Lanczos method is a diagonalization scheme that yields $m$ approximate eigenvalues and -vectors of a given matrix, where typically $m \ll D$ holds. The biorthogonal version  \cite{SaadSparseLinearSystems2003} is suited especially for non-hermitian matrices. 

In the conventional Lanczos method the so-called Krylov subspace $\mathcal{K}_m = \{ R^n \vec{\phi}_0, n \in [0,m-1] \}$
is generated by choosing a starting vector $\vec{\phi}_0$. Then, this Krylov subspace  is orthogonalized by a Gram-Schmidt algorithm. By this procedure, an $m \times m$ tridiagonal matrix $T_m$ can be generated iteratively. From the eigenvalues and eigenvectors of $T_m$ the corresponding Ritz values/vectors of the original matrix can be computed.

For a nonsymmetric matrix $R$, a corresponding left Krylov subspace $\mathcal{K}^L_m = \{ \vec{ \phi_n}= [R^h]^n \vec{\phi}_0, n \in [0,m-1] \}$ needs to be constructed and orthogonalization is performed between states of the left and the right space similar to
co- and contravariant vectors in non-orthogonal spaces. For further details on the algorithm the reader is referred to Yousef Saad's book \cite{SaadSparseLinearSystems2003} on iterative methods for sparse linear systems.

$\vec{\phi}_0 = \vec{\rho}(t=0)$ is chosen as a left starting vector $\vec{w}_0$ as well as a right starting vector $\vec{v}_0$ for the Lanczos method while one of them needs to be normed by $1/ \langle \vec{w}_0,\vec{v}_0 \rangle $. This choice yields an accurate short-time solution for Eq.\ \eqref{eq:LindbladED} which can be understood by first expanding $e^{-Rt}$ into a Taylor series before inserting a complete eigenbasis $\sum_{k=1}^m \ket{v_k} \bra{w_k}$ spanning the Krylov subspace.  The overlap matrix elements $c_{k} = \langle \vec{w}_k \ket{\rho(t=0)}$
and the approximate eigenvalues $\lambda_k$ obtained by the Lanczos method  enter the Taylor expansion
\begin{eqnarray}
\vec{\rho}(t) &=& \sum_{k=1}^m \sum_{n=0}^{m-1}  \lambda_k^n \ket{v_k} c_{k} \frac{(-t)^n}{n!} + \mathcal{O}(t^{m}) \komma
\end{eqnarray}
indicating that the accuracy increases with increasing Krylov subspace dimension $m$.

\subsubsection{The eigenspectrum of the Bloch-Redfield tensor}

Since the Bloch-Redfield tensor $R$ in Eq.\ \eqref{eq:BRT} is nonsymmetric, the spectrum of eigenvalues $\lambda_k$ is generally complex. The master equation  for the DDM, however, ensures that the eigenvalues as well as the eigenvectors are real to maintain the hermitian 
property of the total density matrix. For the ODDM, all complex values can be ordered in complex-conjugated pairs. 

The Lanczos approach, however, can also be used in our context to make very accurate predictions
on the long-time behavior. In general, the true eigenvalues $\lambda_k$ of the tensor $R$ for the ODDM 
are finite and $\Re \lambda_k > 0$. If the Lanczos approach maintains the condition $\Re \lambda_k > 0$ even the approximative solution in a reduced $m\times m$ space yields a complete decay of the ODDM with possibly slightly  modified relaxation-time scales.

As discussed above, the  tensor $R$ for the DDM 
has one eigenvalue $\lambda_0=0$ with the  corresponding right steady-state eigenvector $\vec{v}_0$.
Thus, the steady-state density matrix $\vec{\rho}(t \rightarrow \infty) = c_0 \vec{v}_0$ is obtained by calculating the overlap
between the left eigenvector $\vec{w}_0$ and  the initial vector $c_0 = \langle \vec{w}_0 , \vec{\rho}(t=0) \rangle$.
As we have shown  in Sec.\ \ref{sec:occupation-dynamics}, this steady-state density matrix 
obtained via Eq.\ \eqref{DiagPart} which 
is given by the Boltzmann distribution for a system approaching the thermal equilibrium.
As long as this thermal density matrix $c_0 \vec{v}_0$ 
has a finite overlap with the initial density matrix, $c_0\not = 0$,
this vector is always included in the Krylov subspace by construction.

We note that the correct solution for  $\vec{v}_0$ with an eigenvalue $\lambda_0=0$ is always found 
with high precision by the Lanczos approach since it is an extreme eigenvalue.
Therefore, the approximation for the DDM
\begin{eqnarray}
\vec{\rho}(t) &=& e^{- R t} \vec{\rho}(t=0) \approx  \sum_{k=1}^{m} c_k e^{- \lambda_k t} \vec{v}_k,
\label{eq:Lindblad-Lanczos}
\end{eqnarray}
using the Lanczos eigenvectors $\vec{v}_k,\vec{w}_k$ and eigenvalues $\lambda_k$ includes
the correct limit for $t\to\infty$. This reflects the fact that only the very large and the very small (i.e.\ extreme) eigenvalues in  the  Lanczos eigenvalue spectrum \cite{Kuijaars2000} are reliable representations of the true spectrum of a matrix. Therefore, the Lanczos approach has been successfully used for the calculation of ground states of finite size Hamiltonians.

\section{Benchmark}
\label{sec:benchmark}

\subsection{The resonant level model}
\label{subsec:RLM}

Since the exact  solution of the local dynamics in the resonant level model (RLM) is known \cite{AndersSchiller2006}
we will use it to benchmark our hybrid NRG approach. Throughout this paper, a symmetric box density of states $\rho(\epsilon) = \rho_0 \Theta(D - |\epsilon|)$ is used in all TD-NRG calculations.

The Hamiltonian of the RLM describes the hybridization of a localized level at the energy $E_d$  with a conduction band 
\begin{align}
H = E_d(t) d^\dag d + \sum_k \epsilon_k c_k^\dag c_k^{ } +  V \sum_k \{ d^\dag c_k^{ } +c_k^\dag d \}
\label{eq:H_RLM}
\end{align}
where $c_k^\dag$ creates a spinless conduction electron with momentum $k$ and energy $\epsilon_k$
and $d^\dagger$ creates an electron on the localized level. We also allow for a time dependency of the single-particle
energy $E_d(t)$. Here $\Gamma_0 = \pi \rho_0 V^2$ is the hybridization width and $\rho_0$ is the
conduction-electron density of states at the Fermi energy.

To adapt the RLM to our hybrid approach, we set $H_{\rm imp} = E_d(t) d^\dag d$ 
in  Eq.\ \eqref{eq:first_H}, and $H_{I}$ is given by Eq.\ \eqref{eqn:Hhyp}.
Since the number of bath flavors $M=1$, we drop the index $\nu$ in the following.

\subsection{Real time dynamics}
\label{Sec:small-chains-vs-hybrid-TDNRG}

Choosing $\hat{n}_d = d^\dag d$ as the observable $\hat{O}$ in Eq.\ \eqref{eqn:hybrid-time-evolution-intro}, we consider a stepwise change in the energy of the level: $E_d(t) = \Theta(-t) E_d^{i} + \Theta(t) E_d^{f}$. In the wide-band limit ($D\gg \Gamma_0$) $n_d(t) = \langle \hat{n}_d(t) \rangle$ can be solved exactly in closed analytical form using the Keldysh formalism \cite{AndersSchiller2006}. 
For $T=0$, the analytic solution features an exponential decay from the initial equilibrium occupancy of $\mathcal{H}_i$ to the new equilibrium occupancy of $\mathcal{H}_f$ with two decay rates $\Gamma_0$ and $2 \Gamma_0$.

\begin{figure}[t]

(a) \includegraphics[width=0.45\textwidth]{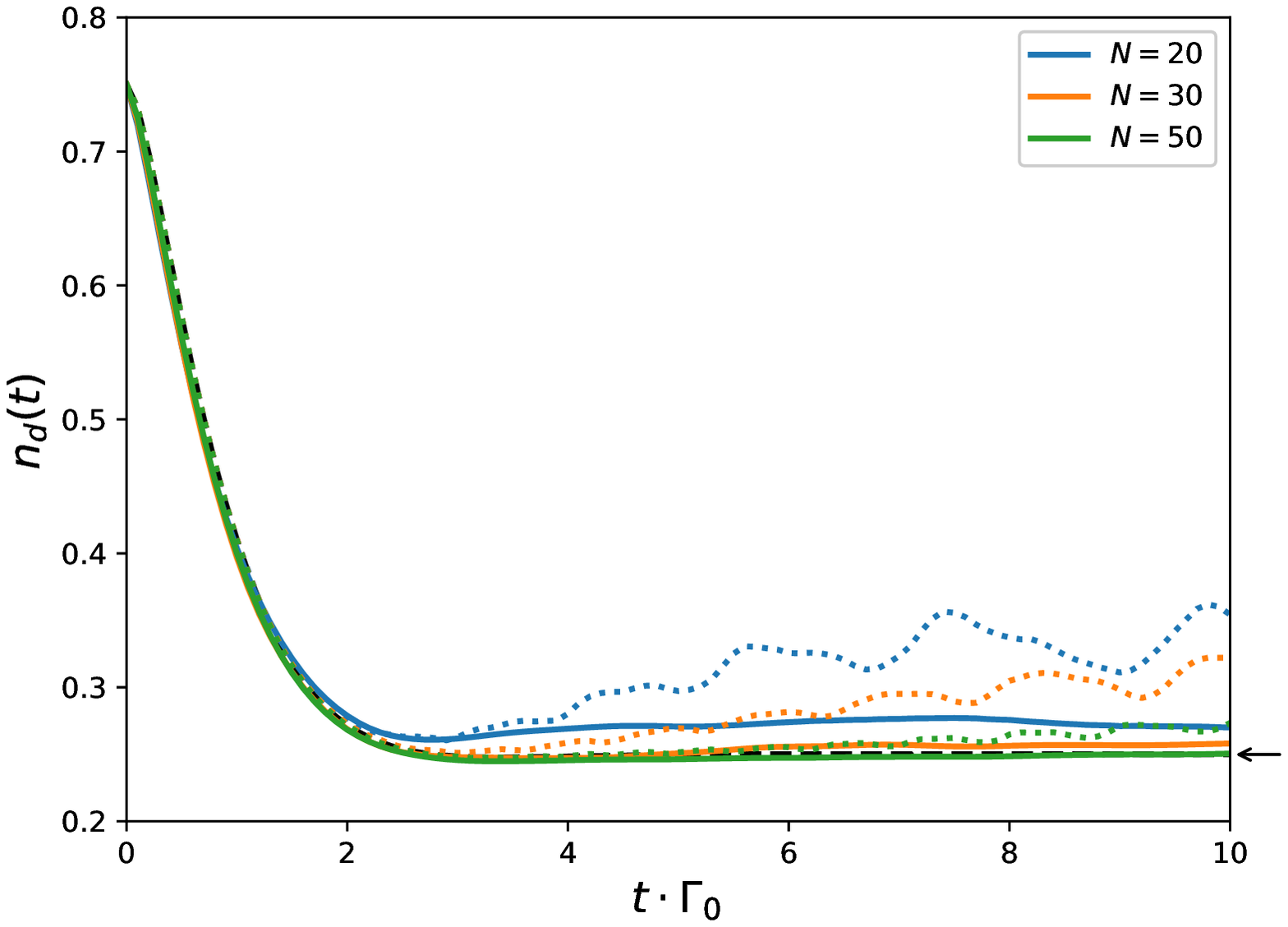}

(b)  \includegraphics[width=0.45\textwidth]{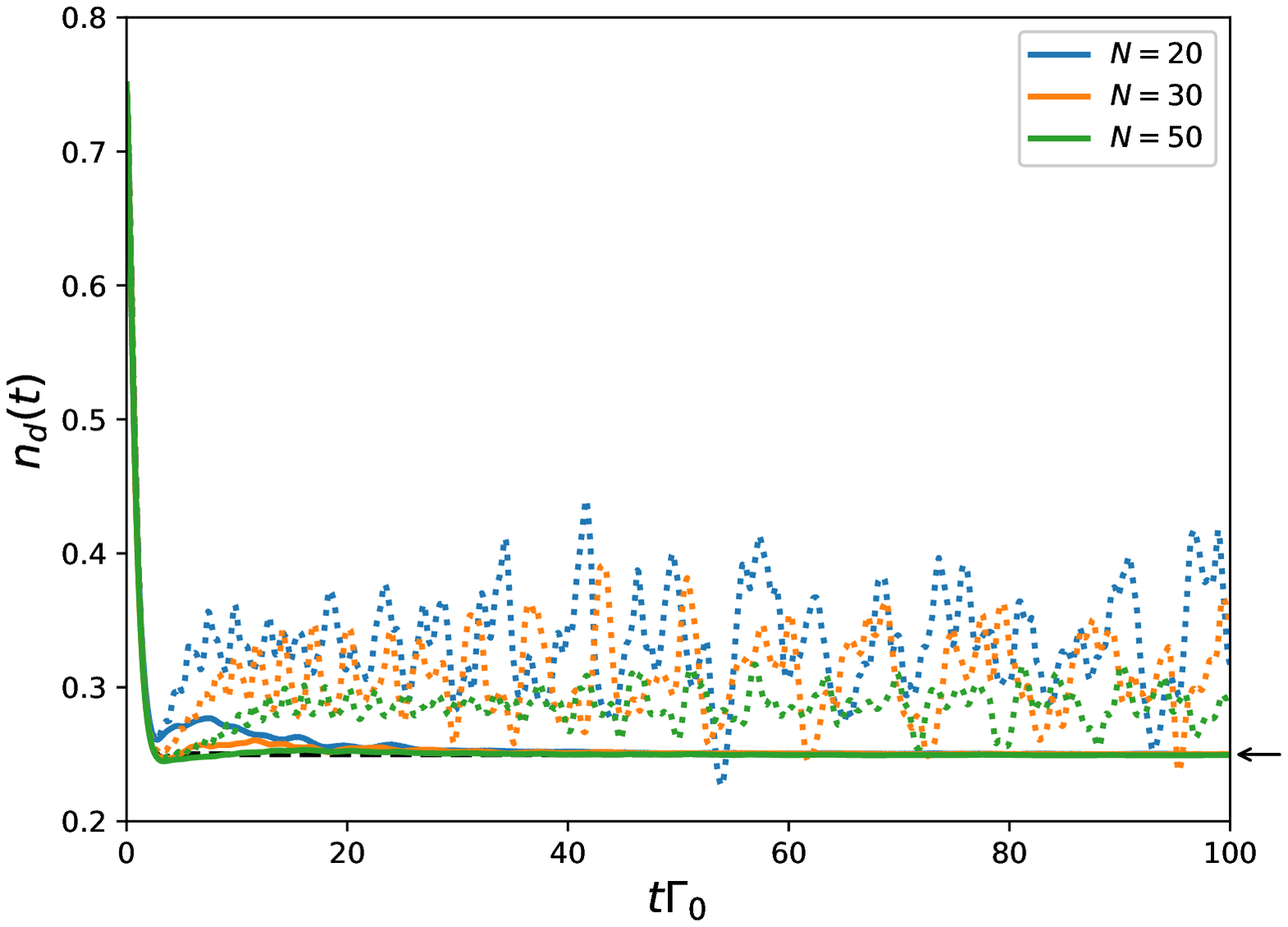}
 
\caption{
Real-time dynamics of the local orbital occupancy  $n_{\rm d}(t)$ (a) for a short time scale and (b) for long time scales
obtained by the open chain  approach and shown as full lines compared to the  TD-NRG  approach which is added as a dotted line in the same color for the RLM. The exact analytical solution  of Ref.\ \cite{AndersSchiller2006} has been added as black dotted lines. Data for different Wilson chain lengths $N$ is presented for a sudden change in the energy of the level from $E_d^{i} = -\Gamma_0$ to $E_d^{f} = \Gamma_0$. The NRG parameter $\Lambda$ ($\Lambda=1.59, N=50; \Lambda=2.17, N=30; \Lambda=3.21,N=20$) was adjusted such that the same temperature $T=0.01\Gamma_0$ is reached for all curves; the corresponding numbers of NRG iterations $N$ are stated in the legend. NRG parameters used: $ D=10^3 \Gamma_0, N_{\rm S} = 10^3, N_z=4$.}

\label{fig:RLMGuettge}
\end{figure}

We present data for a sudden level quench in the RLM that leads to a depletion of charge on the impurity in  Fig.\ \ref{fig:RLMGuettge}.
The real-time dynamics of the local orbital occupancy  $n_{\rm d}(t)$ 
(solid lines) are obtained with our hybrid open chain (OC) approach, 
Eq.\ \eqref{eqn:hybrid-time-evolution-intro}: 
The constant reduced density matrix $\rho_{s,r}^{\rm red}(m)$
was made time-dependent, and its dynamics was calculated by the Bloch-Redfield master equations.
The master equations were solved via a biorthogonal Lanczos algorithm \cite{SaadSparseLinearSystems2003}. 
The dimension of the Krylov subspace for calculating the real-time dynamics of 
the diagonal matrix elements was set to $m=1000$, while  a Krylov subspace dimension of $m=100$
turned out to be sufficient for obtaining the dynamics of off-diagonal matrix elements that only
require coupling matrix elements within a single Wilson shell. We also supplied the exact analytic solution \cite{AndersSchiller2006} as a black dashed line to the panels.
$n_{\rm d}(t)$ was calculated for three different Wilson chain lengths $N$ by varying the NRG parameter $\Lambda$ to ensure  the same target temperature $T=0.01\Gamma_0$

For comparison, we added the results obtained by the closed chain (CC)  
TD-NRG approach \cite{AndersSchiller2006,AndersSchiller2006} for the same
parameters as dotted lines of the same color. The arrow marks the thermodynamic expectation value of the equilibrium NRG using the finial Hamiltonian $\mathcal{H}_f$. We  z-averaged the dynamics 
using $N_z=4$ different NRG chain representations \cite{YoshidaWithakerOliveira1990,AndersSchiller2005,AndersSchiller2006}.
The z-averaging significantly  reduces the finite size oscillations, but  the charge occupation in the CC results still does not converge to the thermodynamic limit as expected from the exact continuum limit. 

The NRG and the quench parameters are chosen close to Fig.\ 1(a) of Ref.\ \cite{EidelsteinGuettgeSchillerAnders2012} in order to make a connection to the literature. 
Usually, the  averaged TD-NRG steady-state long-time limit is close to the thermodynamic NRG expectation value. These quench parameters, however, are deliberately chosen such that the deviation is large due to back reflections along the NRG chain as discussed in Ref.\ \cite{EidelsteinGuettgeSchillerAnders2012}.

For short-time scales, the TD-NRG and our OC approach track the exact result very accurately. The differences between the approaches become pronounced in the long-time limit plotted in  Fig.\ \ref{fig:RLMGuettge}(b)
 illustrating the influence of the NRG parameter $\Lambda$ onto  the real-time dynamics.
It is well understood \cite{Schmitteckert2010,EidelsteinGuettgeSchillerAnders2012} that the exponential decay of the tight-binding parameters of the Wilson chain leads to a tsunami effect \cite{Schmitteckert2010} of a severe slowdown of charge transport along the chain: The charge transport velocity mismatch leads to  back reflections that increase with increasing $\Lambda$ and are the origin 
of the deviation between the calculated real-time dynamics and the exact analytical solution. 
This problem is solved by  including the additional reservoirs perturbatively 
in the dynamics. The thermal  state is reproduced as a steady state in all cases with the largest deviations at intermediate times for the largest value of $\Lambda$. In this case, the TD-NRG shows the largest deviations as well. Furthermore, the bath couplings are the largest in this case so that the second order perturbation theory treating the reservoirs is insufficient to fully reproduced the exact solution. However,
Fig. \ref{fig:RLMGuettge}(b) clearly demonstrates the convergence for $\Lambda\to 1^+$: the choice of $\Lambda=1.59$ already excellently tracks the exact analytic solution for the continuum problem.

The plots in   Fig.\ \ref{fig:RLMGuettge} estabish the very good agreement of our proposed hybrid TD-NRG approach with the exact analytical result in the long time limit. The OC approach provides an efficient mechanism for particle exchange with the additional reservoirs such that charge conservation is maintained in the coupled system but excess charge is balanced by the infinitely large reservoirs that couple to each chain site.

\begin{figure}[t]
\includegraphics[width=0.5\textwidth]{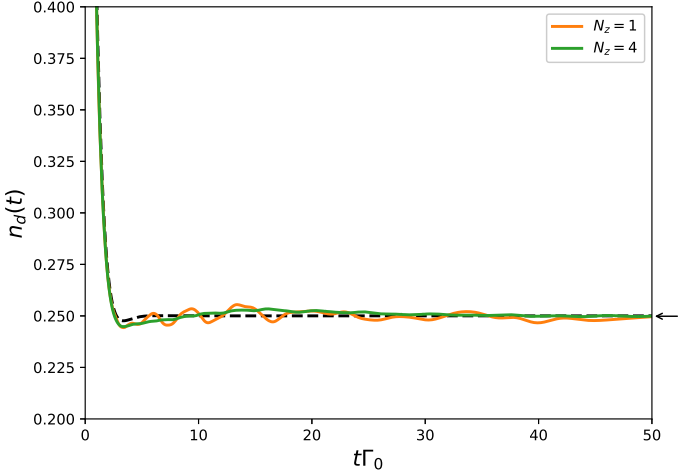}
\caption{
Real-time dynamics of the local orbital occupancy  $n_{\rm d}(t)$  
for $N=50$ with (green) and without (orange) z-averaging \cite{AndersSchiller2006}. All other 
parameters as in Fig.\ \ref{fig:RLMGuettge}.}

\label{fig:RLM-ztrick}
\end{figure}

The effect of the z-averaging \cite{YoshidaWithakerOliveira1990,AndersSchiller2005,AndersSchiller2006} is illustrated in Fig.\ \ref{fig:RLM-ztrick}. The $N_z=4$ result of Fig.\ \ref{fig:RLMGuettge} for the Wilson chain of length $N=50$ (green) is plotted in comparison to the data without z-averaging ($N_z=1$, orange curve). 
The discrepancy between the different data obtained from the OC approach is small: The z-averaging evens out the finite size oscillations which are very close to the exact solutions plotted as a black dashed line. The hybrid approach perfectly reproduces the thermal value of the occupation as indicated by the black arrow at the right sight of the figure and follows the exact solution very accurately.

An important component of our hybrid approach is the coupling of the reduced density matrix elements between all Wilson shells. Since the calculation of all Bloch-Redfield tensor elements are in principle possible -- see Sec. \ref{sec:redfield-diagonal} -- but  numerically too expensive for a practicable implementation, we only calculate the shell diagonal tensor matrix elements and those between adjacent shells $m'=m\pm 1$ in a complete manner. For the coupling of iterations with $|m_1 - m_2| > 1$ we use the approximation $X_{l_1,l_2}^{\tilde m}(m_1,m_2) \rightarrow d^{N- \rm max{(m_1,m_2)}} \delta_{Q_1, Q_2+1}$ as introduced in Sec. \ref{sssec:approx_of_rates}.

\begin{figure}[t]
\includegraphics[width=0.5\textwidth]{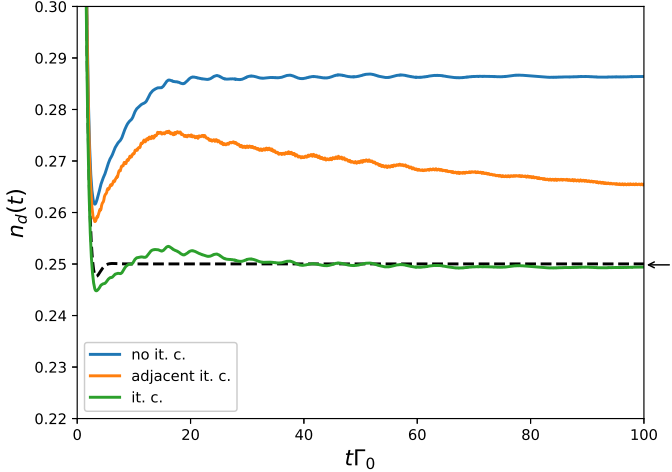}
\caption{
Impurity occupancy $n_{d}(t)$ obtained by the full OC hybrid approach (green),
by an adjacent  approximation for $\Xi_{l_1,l_2}(m_1,m_2)$ including $m_1=m_2$ and
$m_1=m_2\pm 1$ (orange) and restricting to  $m_1=m_2$ (blue). NRG parameters as in Fig.\ \ref{fig:RLMGuettge} for $N=50$.
}
\label{fig:RLM-itc}
\end{figure}

The effect of different approximations to the Bloch-Redfield tensor is depicted in Fig.\ \ref{fig:RLM-itc}. We augmented the  OC approach data for $N=50$ taken from Fig.\  \ref{fig:RLMGuettge} with the results obtained with additional approximations in
calculations of diagonal density matrix elements. 

The blue curve (no shell coupling) is obtained by a tensor $\Xi_{l_1,l_2}(m_1,m_2)$ that is diagonal in the Wilson shell indices, i.\ e.\ $\Xi_{l_1,l_2}(m_1,m_2)=\delta_{m_1,m_2} \Xi_{l_1,l_2}(m_1,m_1)$. The coupling of the additional reservoirs generates a damping in the real-time dynamics of the orbital occupancy $n_d(t)$. Since the sum of the  diagonal density matrix elements remains conserved in each Wilson shell as in the TD-NRG, the steady-state value is very similar to the time averaged TD-NRG value at infinitely long times: the decay into the thermal steady-state is not possible without coupling the discarded states of different iterations $m$.

\begin{figure}[t]
\includegraphics[width=0.5\textwidth]{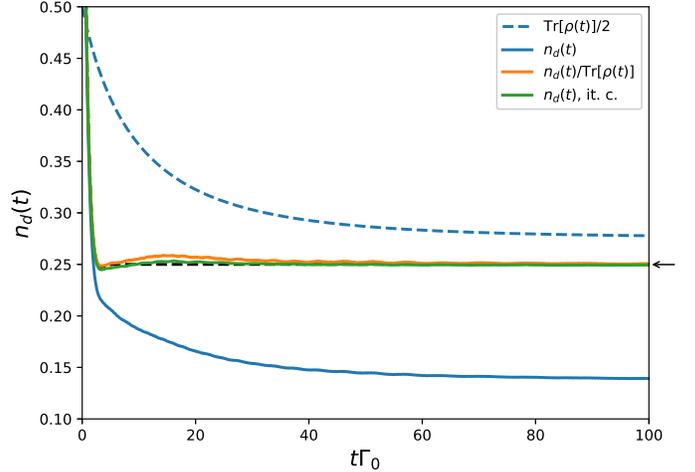}
\caption{
Impurity occupancy $n_{d}(t)$ obtained by the OC hybrid approach but with different approximations.
The green curve is obtained by the full algorithm and is taken from  Fig.\ \ref{fig:RLMGuettge}.
For the blue curve, we neglected the coupling between discarded states of different iterations
and included the relaxation into the kept states for each iteration instead. The loss of the trace of the density matrix as a function of time is shown as a blue dashed line. Normalizing the blue curve by the time dependent trace yields the orange curve. NRG parameters as in Fig.\ \ref{fig:RLMGuettge} for $N=50$.
}
\label{fig:RLM-traceloss}
\end{figure}

This is fundamentally changed when the adjacent approximation which includes all tensor elements $m_1=m_2\pm 1$ is applied (depicted as an orange line). 
We notice a decay of $n_d(t)$  at intermediate times
even though there is no convergence on the time scales plotted in Fig.\ \ref{fig:RLM-itc}. However, we
proved analytically in the appendix \eqref{Eq:ConservTrace} as well as numerically - not shown here - that the thermal expectation value of $n_d$ with respect to $H_f$ already is obtained as the steady-state value in this approximation. The decay rate, however, is very low. This problem is solved by our approximate treatment of all other matrix elements $\Xi_{l_1,l_2}(m_1,m_2)$ that includes a coupling of the diagonal density matrix elements of all Wilson shells with exponentially decaying matrix elements that are allowed by the symmetry but ignoring the precise values of the  overlap matrix elements (green curve).

In Fig.\ \ref{fig:RLM-traceloss} we plot the all-coupling approach (green) versus a complete separation of the iterations (blue), comparable to the blue curve in Fig. \ref{fig:RLM-itc}.
The difference lies in the fact, that we now include all states, discarded and kept, at all iterations for the independent Bloch-Redfield equations. This implies a realistic relaxation of the high energy states into the low energy kept states for each NRG iteration.
Since the kept states of the diagonal part of the density matrix, however, are not included in Eq. \eqref{eqn:hybrid-time-evolution-intro} we end up with an effective unphysical loss of the trace. 
This can easily be compensated for by artificially dividing any non-equilibrium expectation value by the time-dependent trace and thus ensuring to keep the trace of the resulting expression constant (orange curve). That way a correct thermalization can be realized. 
Even though this  approximation is very efficient regarding computation time and memory requirements, its motivation is unphysical. For that reason we will continue this paper by using the approach that couples all iterations and thus includes an inherent conservation of the trace.

\section{Real time dynamics for correlated models using the open chain approach}

After establishing the quality of the OC algorithm to the non-equilibrium dynamics by comparing the results of the approach to the exact analytical solution of the occupancy dynamics in the RLM, we apply our approach to two problems for which an exact analytic solution is unknown: the interacting resonant level model and the single impurity Anderson model.

\subsection{Interacting resonant level model}

In order to proceed to the first non-trivial problem of this paper, the RLM is extended by a Coulomb repulsion $U$ between the local impurity level and the band which defines the interacting resonant level model (IRLM). Here the  modified impurity Hamiltonian  $H_{\rm imp}$  reads
\begin{align}
H_{\rm imp} = E_d(t) d^\dag d + U \left(d^\dag d^{ } - \frac{1}{2}\right) \left(f_0^\dag f_0^{ } - \frac{1}{2} \right) \punkt
\label{eq:H_IRLM}
\end{align}
This model has been intensively studied \cite{VigmanFinkelstein78-1,VigmanFinkelstein78-2,Schlottmann1980} in the 1970s 
due to its connection to the Kondo problem  \cite{Schlottmann1978}.
In resent years, the interest has shifted to its non-equilibrium properties, particularly for a biased two-lead setting \cite{RT-RG-IRLM-2,FRG,MethaAndrei2005,BoulatSaleurSchmitteckert2008}.

The IRLM shares the line of low-energy fixed points with the non-interacting RLM after renormalization of 
\begin{align}
\Gamma_0\to \Gamma_{\rm eff}  \approx D (\Gamma/D)^{1/(1+\alpha)},
\label{eq:Gamma}
\end{align}
with $\alpha = 2 \delta - \delta^2$ and $\delta = (2/\pi) \arctan{(\pi \rho U/2)}$. 
Nevertheless the non-equilibrium dynamics of both models differs significantly  \cite{EidelsteinGuettgeSchillerAnders2012,GuettgeAndersSchiller2013}.
While the coherent oscillations present in the analytic solution \cite{AndersSchiller2006} are strongly damped in the RLM and, therefore, are only observable for extreme parameter choices, an increasing number of coherent oscillations in $n_d(t)$ is found with increasing $U$ \cite{EidelsteinGuettgeSchillerAnders2012,GuettgeAndersSchiller2013}
in the IRLM. The additional Coulomb repulsion $U$ favors the single-electron subspace spanned by the impurity orbital  and the first Wilson chain site. 
The coherent oscillation frequency is given by the energy difference between the binding and anti-binding molecular state formed by the hybridization since the initial configuration can be expanded into these two local states with different eigenenergies. In the limit of large $U$, the rest of the Wilson chain is essentially decoupled from those two states, and the virtual charge fluctuations between these states and the rest of the Wilson chain induces a damping of these coherent oscillations that is proportional to $U^{-2}$  \cite{GuettgeAndersSchiller2013}.

\begin{figure}[t]
 \includegraphics[width=0.5\textwidth]{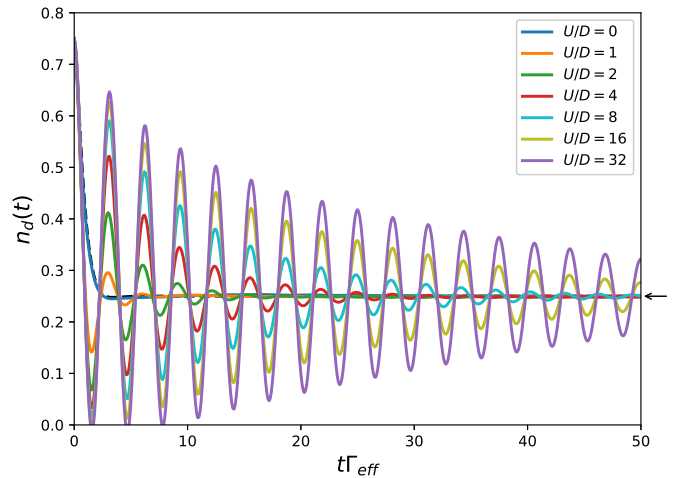}
\caption{
The real-time dynamics of $n_d(t)$ vs time in the IRLM for different values of $U$ obtained by our OC hybrid approach (solid lines) for a sudden change in the energy of the level from $E_d^{i}/\Gamma_{\rm eff} = -1$ to $E_d^{f} /\Gamma_{\rm eff} = 1$. The analytical $U=0$ result is added as a guidance (black dashed line). The thermodynamic expectation value $n_d^f$ is added as a black arrow on the r.h.s of the figure for comparison. NRG parameters: $\Lambda=1.59, N=50, D/\Gamma_{\rm eff}=10^3, N_{\rm S} = 10^3, N_z=4$ so that $T/ \Gamma_{\rm eff}=0.01$.
}
\label{fig:IRLM_all}
\end{figure}

In order to ensure quenches between the same initial and final equilibrium fixed points,
the hybridization strength $\Gamma_0$ has been adjusted such that $n_d(0)=0.75$
and $n_d(\infty)=0.25$ for all values of $U$, implying $E_d^{i}/\Gamma_{\rm eff} =-1$
and $E_d^{f}/\Gamma_{\rm eff} =1$ for all curves.
The OC results for the local occupancy $n_d(t)$ are shown in Fig.\ \ref{fig:IRLM_all}.
Upon increasing $U$ a new time scale $\tau_U$ emerges which is much larger than the thermodynamical relaxation time scale $\tau_0 \propto 1/\Gamma_{\rm eff}$. 
The time scale $\tau_U$ characterizes the decay of the amplitude of coherent oscillations.
For $U \rightarrow \infty$ the charge  simply oscillates between the impurity and the first Wilson chain site, while for a finite $U$ the oscillations are damped and the system approaches thermal
equilibrium.

\begin{figure}[t]
 \includegraphics[width=0.5\textwidth]{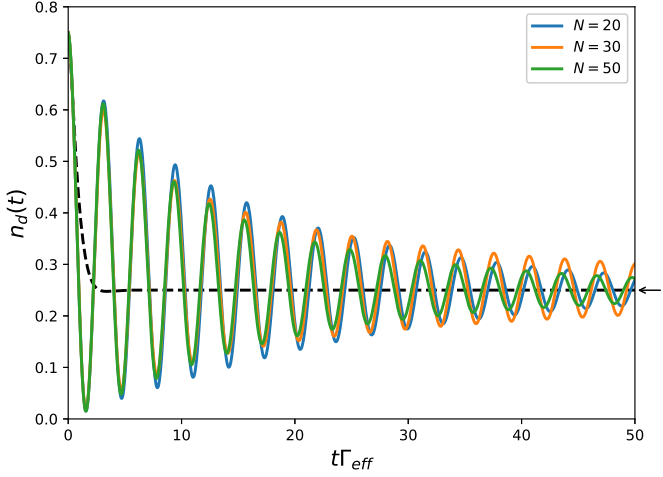}
\caption{
The real-time dynamics of $n_d(t)$ vs time in the IRLM for a fixed $U/D=16$ 
but different chain length $N$ and NRG discretization parameter $\Lambda$ combinations.
The NRG parameter $\Lambda$ ($\Lambda=1.59, N=50; \Lambda=2.17, N=30; \Lambda=3.21,N=20$) was adjusted such that the same temperature $T=0.01\Gamma_{\rm eff}$ is reached for all curves and $N_S=300$. The analytical $U=0$ curve is added as a dashed line for illustration purposes.
}
\label{fig:IRLM_Lambda}
\end{figure}

Since the partitioning of the original continuum
depends on the NRG discretiation parameter $\Lambda$, we investigated 
the non-equilibrium dynamics of $n_d(t)$ for a fixed value of $U/D=16$ and the same local
quench parameters as used in Fig.\ \ref{fig:IRLM_all} but for three different values of
$\Lambda$. The corresponding chain lengths are adjusted such 
that the effective temperature is the same
for all three cases. The results are plotted in Fig.\ \ref{fig:IRLM_Lambda}.
Remarkably little effect of $\Lambda$ on the oscillation frequency and the relaxation time
is found, although $\Lambda$ strongly influences the spectral weight of the coupling to the additional reservoirs. This indicates that our OC approach is rather robust, and the results depend only 
weakly on the discretization parameter.

\begin{figure}[t]
 \includegraphics[width=0.5\textwidth]{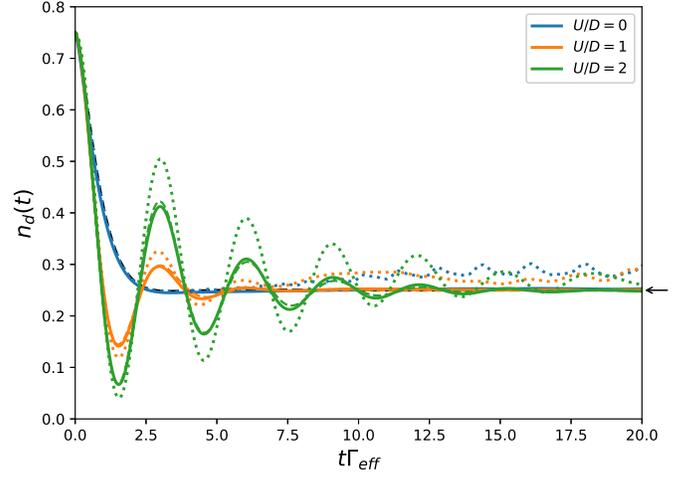}
\caption{
The real-time dynamics of $n_d(t)$ vs time in the IRLM for different values of $U$ obtained by our OC hybrid approach (solid lines) and by the TD-NRG (dotted line) in the same color as well as a fit to Fermi's golden rule (dashed line). NRG parameters as in Fig.\ \ref{fig:RLMGuettge}.
}
\label{fig:IRLM_fit}
\end{figure}

The difference of our approach and the TD-NRG in the IRLM is illustrated for a few small values of $U$ in Fig.\ \ref{fig:IRLM_fit}. Although the oscillation frequency is the same as reported by Guettge {\it et al.}\ \cite{GuettgeAndersSchiller2013}, we note that the decay time  $\tau_U$ of the OC approach is shorter than predicted by the CC approach. The analytical  golden rule estimate  of Ref.\ \cite{GuettgeAndersSchiller2013} is based on a closed chain topology where the impurity orbital and the first Wilson chain site ($m=0$) only couple via the hopping parameter $t_0$ to the rest of the system. The Fermi's golden rule calculation treats the first two orbitals as a closed system and adds a perturbative coupling to the rest of the chain.  The long-time artefacts of the CC approach are suppressed in Ref.\ \cite{GuettgeAndersSchiller2013} by combining the TD-NRG with a TD-DMRG 
approach for a very long tight-binding chain and stopping the simulation before reflections
at the chain end are detectable at the impurity. In our approach, the additional reservoirs cause an additional decay of the coherent oscillations and ensure the thermalization to the expectation value.

\begin{figure}[t]
 \includegraphics[width=0.5\textwidth]{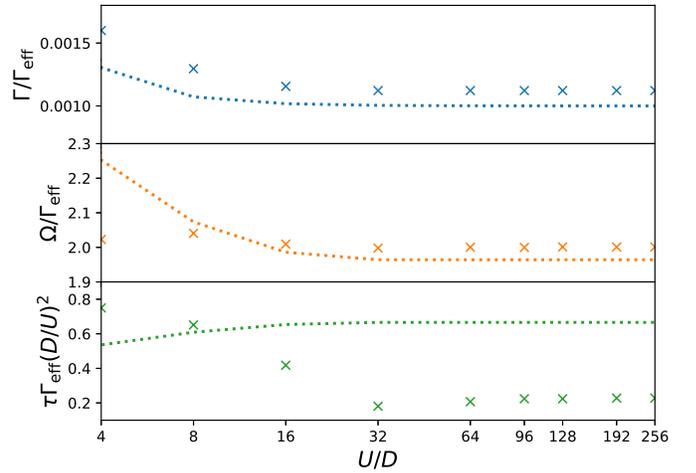}
\caption{
A comparison of analytical estimates (dotted lines) and the numerical values (stars) for three different IRLM parameters. The coupling strength $U/D$ has been varied.}
\label{fig:IRLM_params}
\end{figure}

In  Fig.\ \ref{fig:IRLM_params}, we present a comparison of numerically extracted parameters
with their analytical predictions. In the top panel, we show the NRG results for the ratio
$\Gamma_0/\Gamma_{\rm eff}$ as a solid line,  $\Gamma_0$ being the bare hybridization strength of the model. The results of the perturbative RG prediction according to  Eq.\ \eqref{eq:Gamma} have been added as a dotted line. Both graphs agree excellently in the limit of large $U$. The middle panel and the bottom panel of Fig.\ \ref{fig:IRLM_params} present the numerical fit to the analytical golden rule results stated in Eq.\ (17) of Ref.\ \cite{GuettgeAndersSchiller2013} and their analytical predictions. The oscillation frequency $\Omega$ of the occupation was calculated by $\Omega = \e_+ - \e_- = 2 \sqrt{\left( E_d/2\right)^2 + \left(V_{\rm eff}  \right)^2}$ with $V_{\rm eff}$ being the renormalized hybridization strength parameterizing $\Gamma_{\rm eff} = \pi V_{\rm eff}^2/2D$.

As expected, the analytical prediction agrees very well with the numerical value for the large $U$ regime where the golden rule result is applicable.
Nevertheless, a significant deviation  between the analytical and the numerically extracted
relaxation time $\tau$ is observed. The analytical solution presented in Ref.\ \cite{GuettgeAndersSchiller2013} 
predicts
\begin{eqnarray}
\tau \Gamma_{\rm eff} \left(\frac{D}{U}\right)^2 &=& \frac{\pi^4}{256}\frac{\Omega}{D}\frac{\Gamma_{\rm eff}}{\Gamma}
\end{eqnarray}
and is plotted as a dotted green line in the bottom panel of Fig.\ \ref{fig:IRLM_params}.
As mentioned above regarding Fig.\ \ref{fig:IRLM_fit} the relaxation time extracted for the open chain in the IRLM does not exactly match the golden rule prediction. In fact, we approximately obtain an overlay of three different decay times, the smallest one stemming from the DDM. The remaining two decay times damp the oscillations as exponential functions in Eq.\ \eqref{eqn:hybrid-time-evolution-intro}. The largest decay time influences the long time behaviour of the occupation and thus we have chosen this value to be plotted in comparison to the golden rule approximation for $\tau$ in Fig.\ \ref{fig:IRLM_params}. Obviously, the long time relaxation $\tau \propto (U/D)^2$, as predicted in the golden rule, whereas the asymptotic value for large $U$ is smaller, thus implying a faster relaxation, as discussed above. 

In the OC approach, presented here, the fundamental difference to the CC approach is the direct coupling of an auxiliary reservoir to the first Wilson site $m=0$ as well: even if we artificially decouple
the rest of the Wilson chain from the first site by setting $t_0=0$, the oscillations remain damped for any finite $U$ due to the relaxation channel provided by the first bath. In the limit of large $U$ we expect a superposition of two damping channels: damping by the rest chain and damping by the  high-energy modes of the reservoir $\Delta_0(z)$.  This additional damping mechanism in our OC explains the decrease of $\tau_U$ compared to the CC approach as demonstrated in 
Fig.\ \ref{fig:IRLM_fit}. Our OC also avoids the reflections of charge waves propagating along the Wilson chain since they are damped by the reservoirs as expected from the continuum problem.
Furthermore, the analysis of the RLM has already shown that the relaxation times of our approach are slightly exaggerated for $t \Gamma_0 < 10$ (see e.g. Fig.\ \ref{fig:RLM-itc}) which stems from the approximation in Sec.\ \ref{sssec:approx_of_rates} where the matrix elements of the BRT for $|m-m'|> 1$ are still assumed slightly too large. This yields a faster relaxation for short times fading into a smaller rate for later times.

\subsection{Single impurity Anderson model}

\subsubsection{Definition of the model}

In the single impurity Anderson model (SIAM) the  spin degree of freedom $\nu = \sigma$, the onsite repulsion $U$ and an optional local magnetic field strength $b(t)$ are added to the RLM. The SIAM impurity Hamiltonian now reads:
\begin{align}
H_{\rm imp} = \sum_\sigma \left[ E_d(t) - \frac{\sigma}{2} b(t) \right]d^\dag_\sigma d^{ }_\sigma + U d^\dag_\uparrow d^{ }_\uparrow d^\dag_\downarrow d^{ }_\downarrow \punkt
\label{eq:H_SIAM}
\end{align}
We choose the spin quantization axis parallel to the external magnetic field direction and absorb the prefactor $g\mu_{B}$ into the magnetic field strenght $b$ which is consequently measured in the units of energy. Since we are not interested in the limit of large magnetic fields of the order of the band width \cite{KondoModelComparisonGebhard2020}, we neglect the small corrections due to the spin polarization of the conduction band and only apply a local magnetic field for simplicity.
The bath Hamiltonian and the interaction Hamiltonian are given by Eq.\ \eqref{eqn:hybrid-nrg-hamiltionian-N-reservoirs} and Eq.\ \eqref{eq:H_I_N} respectively,  where the spin index $\sigma$ is summed over $M=2$ values.

\subsubsection{Real time spin  and charge dynamics}

\begin{figure}[t]
 \includegraphics[width=0.5\textwidth]{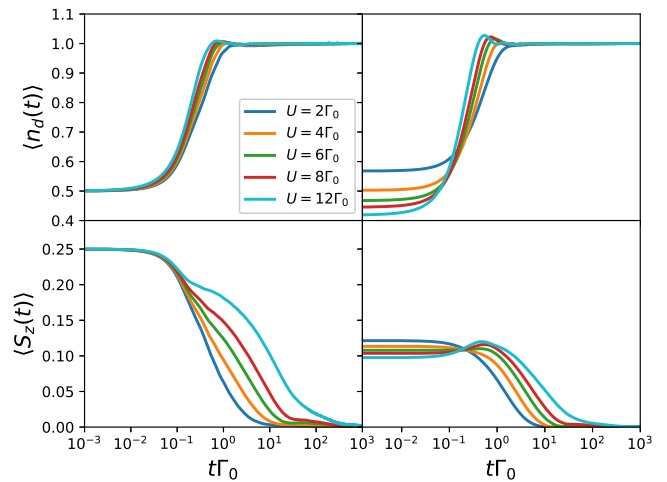}
\caption{
Impurity occupancy $n_d(t)$ and spin polarisation $S_z(t)$ vs time after the quench on a logarithmic time scale. The upper and lower right panel show the results of scenario (i) leaving 
the hybridization strength constant. In the  upper and lower left panel the data after switching the hybridization strength on at $t=0$ are plotted.
Parameters: $\Lambda =1.66, D=20 \Gamma_{0}, N=30, T=0.01 \Gamma_{0}, N_z=4, N_S=10^3$.
}
\label{fig:SIAM_all}
\end{figure}

We apply an instantaneous quench by a change of the parameters $E_d^i \rightarrow E_d^f, 
b_i \rightarrow b_f$, and $\Gamma_i \rightarrow \Gamma_f=\Gamma_0$ at $t=0$. Since the hybridization strength $\Gamma_f=\Gamma_0$ is the same in all cases, all energies are given in units of $\Gamma_0.$ 

We investigated two different quench scenarios: we either (i) keep the impurity hybridization constant, i.\ e.\ $\Gamma_i= \Gamma_f$ or (ii) we switch on  the hybridization at $t=0$. The initial low energy fixed points of both scenarios are fundamentally different. The first case corresponds to the conventional low-energy fixed points of the SIAM \cite{BullaCostiPruschke2008} for the parameter choice of $U$, $E_d$ and $b$, while in the second scenario we start from
 the unstable local moment fixed point where the impurity is decoupled from the conduction band continuum.

In both cases, we leave $U$ constant and only quench $E_d$ and the magnetic field $b$. Initially, we set $b_i=\Gamma_0$ to induce a spin polarisation and switch off the magnetic field at $t=0$. We also start with a degeneracy of the spin-up impurity state and the unoccupied state by setting $E_d^i -b_i/2=0$. For scenario (ii)  the spin-down state is initially completely depopulated so that  $n_d(0)=0.5$, and the local spin polarisation $S_z$ is fixed to $S_z(0)=1/4$. For scenario (i) the initial occupation and spin polarization depend on the ratio $U/\Gamma_0$.

At $t=0$, we quench the level position to $E_d^f=-U/2$ and switch off the magnetic field, $b_f=0$.
Therefore, the thermodynamic low energy fixed point of $H_f$ is the same for all values of $U$ and both scenarios: the particle-hole symmetric strong coupling fixed point.

In Fig.\ \ref{fig:SIAM_all} the dynamics of the impurity occupancy 
$n_d(t)$,
\begin{equation*}
n_d(t) = \langle d^\dag_\uparrow d^{ }_\uparrow  + d^\dag_\downarrow d^{ }_\downarrow \rangle (t)  \komma
\end{equation*}
and the dynamics of the spin polarisation $S_z(t)$,
\begin{equation*}
S_z(t) = \frac{1}{2} \langle d^\dag_\uparrow d^{ }_\uparrow  - d^\dag_\downarrow d^{ }_\downarrow \rangle (t) \komma
\end{equation*}
are plotted as a function of time. The data for five different values of $U$ are shown using our hybrid OC approach. 

\begin{figure}[t]
 \includegraphics[width=0.5\textwidth]{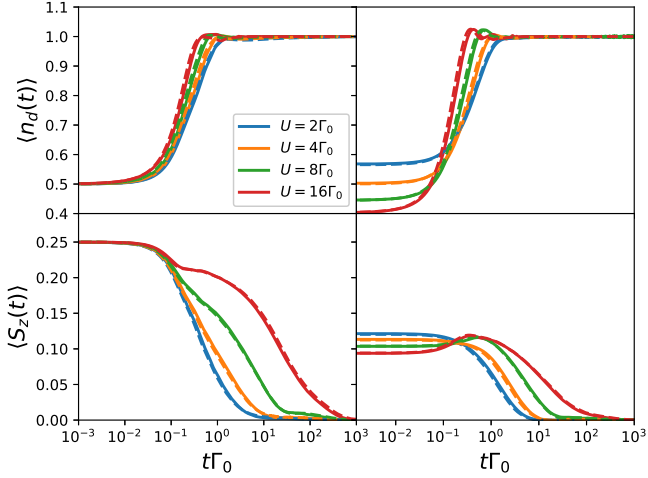}
\caption{
Comparison of $N_S=300$ (dashed lines) to $N_S=1000$ (solid lines) different states for the SIAM regarding the impurity occupation $n_d(t)$ and the spin polarization $S_z(t)$.
}
\label{fig:SIAM_diffS}
\end{figure}

Since the number of states increases by a factor of four in each NRG iteration, 3/4 of the states
are discarded at the end of each iteration in the NRG algorithm. Hence, the number of matrix elements
of the Bloch-Redfield tensors is substantially larger than in the RLM case and the numerical costs of the Lanczos approach for coupling the diagonal density matrix elements become very high. 
While the standard TD-NRG requires around two minutes on todays desktop computers, the OC approach for each of the curves presented in Fig.\ \ref{fig:SIAM_all} took about 3 days on a workstation node utilizing all 16 cores. 

The effect of choosing different numbers of kept states $N_S$ after each NRG iteration is demonstrated
for the SIAM in Fig.\ \ref{fig:SIAM_diffS}. We supplement the data for Fig.\ \ref{fig:SIAM_all} shown as solid lines with $N_S=300$ date (dashed lines) for the same quench parameters. Obviously the differences are very small suggesting the choice of $N_S=1000$ states to be perfectly sufficient for our purpose.

\begin{figure}[t]
 \includegraphics[width=0.5\textwidth]{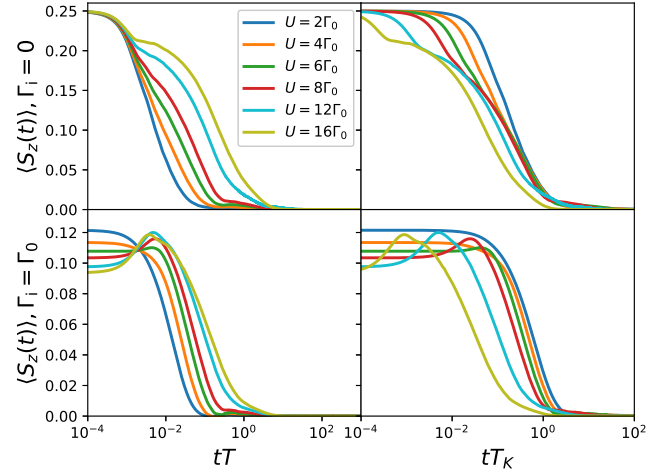}
\caption{
$S_z(t)$ data taken from Fig.\ \ref{fig:SIAM_all} for $\Gamma_i=0$ at the top and $\Gamma_i=\Gamma_0$ at the bottom. The time is scaled by the system temperature $T$ on the left and by the respective Kondo temperature $T_K$ (which depends on $U$) on the right.
}
\label{fig:SIAM_sz-vst-tk}
\end{figure}

The charge relaxation and the spin relaxation occur on different time scales \cite{AndersSchiller2005}
as can already be seen in Fig.\ \ref{fig:SIAM_all}. While the charge relaxation occurs on the scale
set by $\Gamma_f=\Gamma_0$, the spin decay time shows significant  $U$-dependency. 
The equilibrium energy scale that governs the crossover from the local moment fixed point into the strong coupling fixed point is the Kondo temperature $T_K$. This parameter is a measure for the temperature at which the local magnetic moment is already 70\% screened \cite{Wilson75}.

In order to investigate the spin dynamics in more detail, we plotted the $S_z(t)$ data shown in Fig.\ \ref{fig:SIAM_all} versus the dimensionless times $t T$ and $t T_K$ in Fig.\ \ref{fig:SIAM_sz-vst-tk}, where $T$ is the system temperature. For scenario (ii) - top right panel - we find a very good universality  of the long-time behavior of $S_z(t)$. This scenario starts from the local moment fixed point with
a decoupled impurity and approaches the symmetric strong coupling fixed point and, therefore, partially tracks a thermodynamic flow. The dynamics is clearly governed by the Kondo scale for large Kondo temperatures where $T\ll T_K$. The long-time tails of the $U/\Gamma_0=2,4,5,8$ curves show universality. Since $T_K(U/\Gamma_0=8)=0.046 \Gamma_0$, we start to see deviations since the temperature is $T/\Gamma_0=0.01$ in all simulations. For $U=16\Gamma_0$ the system temperature $T\approx 4.2T_K$ is clearly above the Kondo temperature. The top left panel of Fig.\ \ref{fig:SIAM_all} suggests that the relevant decay scale is set by the thermal fluctuations for $T> T_K$, as $S_z(t)$ decays on the scale of $1/T$.

For the scenario (i) depicted in the two lower panels of  Fig.\ \ref{fig:SIAM_all}, the Kondo temperature does not provide such an universal scaling. The characteristic decay time is of the order of $T_K$ for temperatures $T\ll T_K$ but it depends on the initial preparation of the system. Upon increasing the relative temperature $T/T_K$, the thermal fluctuations start to dominate the decay time as in scenario (ii).

As a further indication for the correctness of the TD-NRG results an analytic solution will be used for case (i). The dynamics of the density operator is calculated up to second order in the impurity coupling function.This solution is only valid on short time scales and becomes asymptotically exact in the limit $t \to 0$. The calculation requires a numerical evaluation at finite temperature but
in the limit of $T \rightarrow 0$ we arrive at the compact analytical expression
\begin{align}
n_d(t) &= \frac{1}{2} + 2 B(t,U/2) \non
S_z(t) &= \frac{1}{4} -  B(t,-U/2)
\label{eq:Siam_ana}
\end{align}
with 
\begin{align}
B(t,\e) = \frac{\Gamma_0 t}{\pi} \left. \text{Si}(\w t) + \frac{\Gamma_0}{\w}\frac{\cos(\w t)-1}{\pi} \right|_{\w=-\e}^{D-\e}
\end{align}
and Si$(\e)$ being the sine integral. The full calculations can be found 
in App.\  \ref{app:Analy_siam}.

The OC (solid line) and CC (dotted line)  numerical data for the change of the time-dependent spin (orange) and charge  (blue) expectation values are compared to the analytical curves  (dashed lines) for $U/\Gamma_0 = 2$ and $D/\Gamma_0 = 20$. 
The CC (TD-NRG) agrees perfectly with the analytics for  times $t \Gamma_0 < 0.1$.
Here, the deviation of the OC solution from both curves is clearly visible, but this effect is exaggerated by the double logarithmic plot. The OC and the CC approach merge on time scales $0.1 < t\Gamma_0$. The initial derivations are a generic feature of the  Bloch-Redfield formalism where short-time quantum correlations are ignored due to the factorisation  under the integral.

\begin{figure}[t]
 \includegraphics[width=0.5\textwidth]{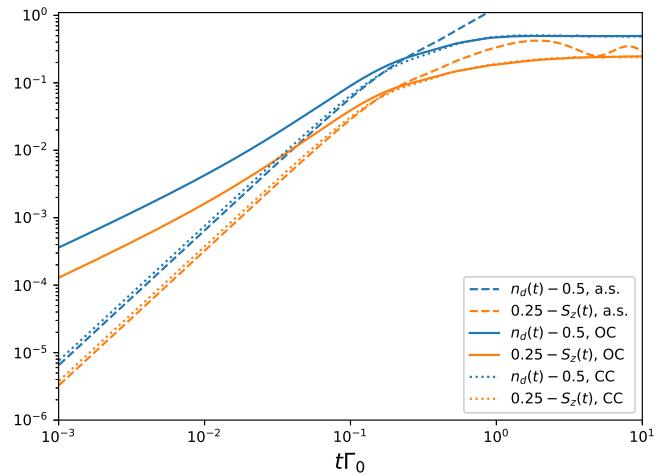}
\caption{
Impurity occupation $n_d(t)$ and spin $S_z(t)$ compared to the closed chain (dotted) and the analytical solution (dashed) according to Eq. \eqref{eq:Siam_ana} plotted on a double logarithmic scale.
}
\label{fig:SIAM_analytical}
\end{figure}

\section{Conclusion}

We presented a hybrid approach to the non-equilibrium dynamics in quantum impurity systems which combines the strength of the NRG and the strength of weak coupling approaches for open quantum systems to restore the original continuum problem. The continuous fraction expansion of the coupling function between the quantum impurity and the environment yields a Hamiltonian representation of the original problem decomposed into a discrete Wilson chain and a set of high-energy additional reservoirs, each  coupled to a single Wilson chain site. These reservoirs  represent the high-energy modes of the original coupling function that only couple indirectly to the quantum impurity. Therefore, the standard NRG is defined as an approximation which neglects the coupling to the additional reservoirs.

A different discretized representation of a quantum impurity system augmented with a Lindblad dynamics was previously considered by Arrigoni and co-workers \cite{DoraArrigioni2015,NussArrigoni2015} in the context of non-equilibrium quantum transport. Their approach treats the Lindblad coupling tensor elements as fitting parameters that are determined by a variational approach. In our method, we are able to analytically construct the exact coupling functions to the additional reservoirs that are required to recover the original continuous coupling function of the problem.

Since the NRG has been established as an excellent tool \cite{BullaCostiPruschke2008} for the equilibrium problem, we propose to augment the TD-NRG with a perturbative Bloch-Redfield treatment of the  coupling to the auxiliary reservoirs. We modified the standard  Bloch-Redfield approach \cite{MayKuehn2000} derived for the full density matrix of a finite size system: The approach is applied to the set of reduced density matrices that are required for the  dynamics  of local observables at and around the quantum impurity in order to handle the huge amount of discarded states generated by the NRG truncation.  The fourth rank Bloch-Redfield tensor is evaluated exactly from the analytically constructed coupling functions to the additional reservoirs within a Wilson shell and for the coupling between the diagonal matrix elements of the density matrix of adjacent shells.
We used the generic scaling properties of the matrix elements to substitute the cumbersome exact enumeration by a simplified analytical form for the tensor elements connecting states of Wilson shells that are far apart from each other. This is justified since the matrix elements decay exponentially with the shell distance $|m_1-m_2|$, and their precise value does not affect the steady state solution of the master equation.

It turns out to be crucial that all diagonal matrix elements of the reduced density matrices of all energy shells are coupled.  We have proven that the steady state of the approach is the NRG thermal equilibrium value for a hybrid system coupled to reservoirs that share a common chemical potential. A different current carrying steady state can be achieved in a two lead setup with different chemical potentials \cite{DoraArrigioni2015,NussArrigoni2015}. This will be subject of a futher publication.

We used the known analytic solution of the RLM \cite{AndersSchiller2006} as a benchmark for the proposed hybrid approach and found an excellent agreement between the analytical and the numerical curves. A comparison of real-time dynamics between the TD-NRG and the open chain hybrid approach was presented for two non-trivial strongly correlated models: the IRLM and the SIAM. In all cases, our hybrid approach significantly reduced the finite size oscillations as well as it removed the slight deviation between the non-equilibrium steady state expectation values and the NRG thermal equilibrium values.

Our hybrid approach has the potential to be extended in two ways: (i) adding leads with different chemical potentials and numerically calculating a current carrying steady state in the strong coupling limit, and (ii) deriving a similar approach for
the NRG spectral functions to remove the necessity for an artificial broadening \cite{BullaCostiPruschke2008}
and replacing it by the physical processes included in the original continuum model prior to the discretization.

\appendix

\section{Relation between the first Wilson-chain parameter $t_0$ and the continuous fraction coupling parameter $V_0$}
\label{app:reservoirs}
Below we will show, that the zeroth reservoir of the open chain, which represents the start of our reservoir algorithm, is sufficient for a Wilson chain parameter $t_0$ of any $\Lambda > 1$. Inserting Eq. \eqref{eqn:gamma-0} into Eq. \eqref{eqn:v0-integral} yields
\begin{eqnarray}
\pi V_0^2 &=& V^2 \int_{-\infty}^{\infty}d\omega
 \frac{{\rm Im} \Delta(\omega)} { {\rm Re} \Delta(\omega)^2 + {\rm Im}\Delta(\omega)^2} 
\\
&=& \frac{2 D}{\pi}\int_{-D}^{D}d\omega \frac{\pi^2 }{ 4\text{artanh}^2(\omega/D)  + \pi^2}  = \frac{\pi}{3} D^2.
 \nonumber 
\end{eqnarray}
Since 
\begin{eqnarray}
t_0^2 &=& \frac{D^2}{4} \frac{(1- \Lambda^{-1})(1+ \Lambda^{-1})^2}{1- \Lambda^{-3}}
\end{eqnarray}
the inequality $V_0> t_0$  follows for any $\Lambda > 1$.

\section{Derivation of the Bloch Redfield approach}
\label{app:bloch-redfield}

The dynamics of the density operator $\rho_I(t)$ 
is governed by the differential equation
\begin{eqnarray}
\partial_t \rho_I(t) &=& i [\rho_I(t), V_I(t)]
\label{VonNeumann-A1}
\end{eqnarray}
in the interaction picture, where the system-reservoir coupling takes the form
\begin{eqnarray}
V_I(t) &=&e^{ iH_0 t} H_{I}(N) e^{-iH_0 t}
\end{eqnarray}
and 
\begin{eqnarray}
\rho_I(t) &=& e^{iH_0 t} \rho(t)  e^{-iH_0 t} \punkt
\end{eqnarray}
Here the operators are transformed by $H_0 = H^{\rm NRG}_N   + H_{\rm res}(N)$.\\
For expectation values of local operators
it is sufficient to know the local density operator
$\rho_S(t) = {\rm Tr}_R[\rho_I(t)]$ 
where we have traced out all the reservoir degrees of freedom. This operator is acting only on the Wilson chain
or system $S$ respectively.

Now Eq.\ \eqref{VonNeumann-A1} can be adapted to derive
a Bloch-Redfield equation for the reduced density matrix $\rho_S(t)$ by integrating the equation
\begin{eqnarray}
 \rho_I(t) &=&  \rho_I(0)  + i \int_0^t d t' [ \rho_I(t'),V_I(t')]
\end{eqnarray}
and substituting the resulting $\rho_I(t)$ back into the differential equation. The expression
\begin{eqnarray}
\label{eq:partial-rho-i}
\partial_t \rho_I(t) &=& i [ \rho_I(0),V_I(t)]  \\
&&- \int_0^t d t' [ [\rho_I(t'), V_I(t')] ,  V_I(t)] \nonumber
\end{eqnarray}
is obtained which is used to derive the dynamics of the local 
density operator 
\begin{eqnarray}
\label{eq:partial-rho-i-red}
\partial_t \rho_S(t) &=&  -\int_0^t d t'   {\rm Tr}_R \left[[ [ \rho_S(t') \rho_R , V_I(t')] ,  V_I(t)] \right]
\non
\end{eqnarray}
after tracing out all  reservoir DOFs. This operator is acting only on the DOF of the Wilson chain.
The first term of the r.h.s of Eq.\ \eqref{eq:partial-rho-i} vanishes due to particle number conservation.

In order to derive the dynamics of the reduced density operator
the weak coupling approximation \cite{MayKuehn2000} is employed
and the full density operator $\rho_I(\tau)\approx \rho_S(\tau) \rho_R $ is factorized, where 
$\rho_R$ denotes the equilibrium density operator of the reservoir which remains unaltered by the coupling to the Wilson chain.

The bath coupling functions $\Gamma_{\nu m} (\e)$ derived in  Sec.~\ref{sec:chain-plus-single-reservoir} enter the expression for the greater and lesser reservoir GF
for each reservoir \cite{HaugKoch2004}. The lesser or particle Green function
\begin{subequations}
\label{eq:reservoir-gfs}
\begin{eqnarray}
G^{<}_{\nu,\tilde m}(t,t') &=& i   |t'_{\nu \tilde m}|^2\trb{\rho_R c_{0\nu \tilde m}^\dagger(t) c_{0\nu \tilde m}(t') }
\nonumber
\\
&=& 
i \int_{-\infty}^{\infty} d\e \frac{\Gamma^{H}_{\nu \tilde m} (\e)}{\pi} f(\e)  e^{i\e \tau}
\nonumber \\
&
=& G^{<}_{\nu,\tilde m}(\tau) =  G^{< *}_{\nu,\tilde m}(-\tau)
\end{eqnarray}
and the greater or hole Green function
\begin{eqnarray}
G^{> }_{\nu,\tilde m}(t,t') &=& -i |t'_{\nu \tilde m}|^2  \trb{\rho_R c_{0\nu \tilde m}(t) c_{0\nu \tilde m}^\dagger(t') }
\nonumber
\\
&=& -i \int_{-\infty}^{\infty} d\e \frac{\Gamma^{H}_{\nu \tilde m} (\e)}{\pi} f(-\e) e^{-i\e \tau}
\nonumber
\\
&=& G^{> }_{\nu,\tilde m}(\tau) =   G^{> *}_{\nu,\tilde m}(-\tau) 
\end{eqnarray}
\end{subequations}
only depend on the time difference $\tau=t-t'$
in equilibrium and fully determine the effect of the reservoirs onto the dynamics on the Wilson chain. 
Their Fourier transformations are defined as
\begin{eqnarray}
G^{> (<)}(\w) &=& \int_{-\infty}^\infty d t e^{-i\w t} G^{> (<) }_{\nu,\tilde m}(t).
\end{eqnarray}

\begin{widetext}
The reduced density operator $\rho_S(t)$ obeys the time-local differential equation
\begin{eqnarray}
\label{eq:dgl-rho-red-m}
\partial_t \rho_S(t) &=& 
 -
 i\sum_{\tilde m=0}^{N} \sum_\nu\int_0^t d\tau
\rho_S (t) \left[ f^\dagger_{\nu \tilde m}(t-\tau) f_{\nu \tilde m} (t) G^{>}_{\nu,\tilde m}(-\tau) - f_{ \nu \tilde m}(t-\tau ) f_{\nu \tilde m}^\dagger (t) G^{<}_{\nu,\tilde m}(-\tau)  \right]
\non
&& + i\sum_{\tilde m=0}^{N} \sum_\nu
\int_0^t d\tau 
\left[ f^\dagger_{\nu \tilde m}(t-\tau) \rho_S  (t)  f_{\nu \tilde m}(t) G^{>}_{\nu,\tilde m}(-\tau) - f_{\nu \tilde m}(t-\tau ) \rho_S  (t)  f_{\nu \tilde m}^\dagger (t) G^{<}_{\nu,\tilde m}(-\tau)  \right]
\non
&& + i\sum_{\tilde m=0}^{N} \sum_\nu
\int_0^t d\tau
\left[ f^\dagger_{\nu \tilde m}(t) \rho_S  (t)  f_{\nu \tilde m}(t-\tau) G^{>}_{\nu,\tilde m}(\tau) - f_{\nu \tilde m}(t ) \rho_S  (t)  f_{\nu \tilde m}^\dagger (t-\tau) G^{<}_{\nu,\tilde m}(\tau)  \right]
\non
&&
 - i\sum_{\tilde m=0}^{N} \sum_\nu\int_0^t d\tau
 \left[ f^\dagger_{\nu \tilde m}(t) f_{\nu \tilde m}(t-\tau) G^{>}_{\nu,\tilde m}(\tau) - f_{\nu \tilde m}(t) f_{\nu \tilde m}^\dagger (t-\tau) G^{<}_{\nu,\tilde m}(\tau)  \right]\rho_S  (t)
\end{eqnarray}
\end{widetext}
after substituting the  explicit form 
of $V_I(t)$ into \eqref{eq:partial-rho-i-red} and making use of  the Markov approximation \cite{MayKuehn2000}:
For fast decaying correlation functions $G^{>}_{\nu,m}(\tau), G^{<}_{\nu,m}(\tau)$ relative to the change of $\rho_S(t)$ one can replace $\rho_S(t-\tau)\to \rho_S(t)$ under the integral, converting the integro-differential
equation into a master equation for $ \rho_S(t)$ and neglecting retardation effects. This approximation is the origin of the deviation between the analytical solution and the OC approach in Fig.\ \ref{fig:SIAM_analytical} for very short times.

By calculating the trace on both sides of \myeqref{eq:dgl-rho-red-m} one obtains 
$\partial_t \Tr{\rho_S(t) } = 0$, since for each reservoir GF a pair of terms can be found which cancel each other out. Thus, the derived differential equation conserves the trace of the density operator at all times.

Conservation of the trace under the restriction $m_2 \in \{m_1-1, m_1, m_1+1\}$ (Eq. \eqref{DiagPart} has been used):
\begin{eqnarray}
\sum_{m_1=m_\text{min}}^{N} \sum_{l_1} \dot{\rho}^\text{red}_{l_1,l_1}(m_1;t)  = \sum_{m_1=m_\text{min}}^{N} \sum_{m_2=m_1-1 \geq m_\text{min}}^{m_1+1 \leq N} \sum_{l_1,l_2} \non
\Big(\Xi_{l_2,l_1}(m_2,m_1) \rho^\text{red}_{l_2,l_2}(m_2;t) - \Xi_{l_1,l_2}(m_1,m_2) \rho^\text{red}_{l_1,l_1}(m_1;t) \Big)
\non
\label{Eq:ConservTrace}
\end{eqnarray}
The two sums are interconvertible, so the trace is conserved.

\section{Analytical solution to  short-time dynamics in the SIAM}
\label{app:Analy_siam}

When Eq. \eqref{eq:partial-rho-i} is integrated over time and then inserted into the time-dependent expectation value of any local operator $O$ we obtain
\begin{align}
\label{eq:exp_value_siam}
\langle O(t) \rangle &= \text{Tr} \{ \rho_0 O^I(t) \}  + \langle O'(t) \rangle\\
\langle O'(t) \rangle& \approx  - \int_{0}^{t} d \tau_1 \int_{0}^{\tau_1} d \tau_2  \text{Tr} \{ \rho_0 \left[H^I(\tau_2),\left[H^I(\tau_1),O^I(t)\right]\right] \}\nonumber
\end{align}
after replacing the full dynamics of the density operator by its initial values in the step from line one to line two. This is asymptotically exact for $t\to 0$ and defines a second order approximation in the impurity bath coupling function.
Here
\begin{align}
\label{eqn:C2}
H^I(\tau) =& \sum_{k,\sigma} V_k \left( c^\dag_{k\sigma}(\tau) d_\sigma^{ }(\tau) +  
c^{ }_{k\sigma}(\tau) d_\sigma^\dag(\tau)\right)
\end{align}
is the term for the interaction of the impurity level and the bath excitations. The operators in the interaction representation read
\begin{align}
d_\sigma(t) &= \ket{0}\bra{\sigma} e^{-i \e_d t} - \sigma \ket{-\sigma}\bra{2} e^{-i (\e_d+U) t}\\
c_{k \sigma}(t) &= c_{k \sigma} e^{-i \e_k t} \komma
\end{align}
where $\ket{0}$ is the vacuum state on the impurity, $\ket{2}$ the double occupied state and $\ket{\sigma}$ accounts for either spin state $\uparrow$ or $\downarrow$. 
The density matrix $\rho_0$ factorizes for the interaction quench. We chose the parameter
$E_d^0 = b_0/2 = \Gamma_0/2$   in the 
Hamiltonian \eqref{eq:H_SIAM} for $t<0$.

Inserting Eq.\ \eqref{eqn:C2} into Eq.\ \eqref{eq:exp_value_siam} and evaluating the double 
commutators using the inital density matrix $\rho_0$,  we obtain
\begin{widetext}
\begin{align}
\braket{O'(t)}  =& \frac{2}{Z}\sum_{k,\sigma} V_k^2 A(\epsilon_k-\epsilon_d,t) \cdot \big[ f(\epsilon_k) e^{-\beta E_0} \left(\braket{\sigma} -\braket{0}\right)+ f(-\epsilon_k) e^{-\beta E_\sigma} \left(\braket{0} - \braket{\sigma} \right) \big]\non
 &+\frac{2}{Z}\sum_{k,\sigma} V_k^2 A(\epsilon_k-\epsilon_d-U,t) \cdot \big[ f(\epsilon_k) e^{-\beta E_\sigma} \left(\braket{2} -\braket{\sigma}\right) + f(-\epsilon_k) e^{-\beta E_2} \left(\braket{\sigma} - \braket{2} \right) \big]
 \label{eq:O_prime}
\end{align}
where we have used the shortcut notations
$\braket{s} = \bra{s} O \ket{s}, s \in \{0,\uparrow, \downarrow, 2 \}$ and
$A(\epsilon,t) = \frac{1 - \cos(\epsilon t)}{\epsilon^2}$.
For a constant hybridization function (see Eq. \eqref{eq:hybrid_func}) and applying the low temperature limit, Eq. \eqref{eq:O_prime} can be transformed to
\begin{align}
O^{'}(t) =& \left(\braket{\uparrow} + \braket{\downarrow} - 2\braket{0}\right)
B_{-D,0}(t,\epsilon_d)
 + \left(\braket{0} - \braket{\uparrow} \right)
  B_{0,D}(t,\epsilon_d)
 + \left( \braket{2}-\braket{\uparrow} \right)
 B_{-D,0}(t,\epsilon_d+U).
\end{align}
The integration can be done in an exact manner with
\begin{align}
B_{a,b}(t,\epsilon') = \frac{\Gamma_0}{\pi} \int_a^b d \epsilon A(\epsilon-\epsilon',t) = \frac{\Gamma_0 t}{\pi} \left. \text{Si}\left((\epsilon-\epsilon')t\right) + \frac{\Gamma_0}{\epsilon-\epsilon'}\frac{\cos((\epsilon-\epsilon')t)-1}{\pi} \right|_{\epsilon=a}^b \komma
\end{align}
where Si$(\epsilon)$ is the sine integral. 

For a non-constant hybridization function $\Gamma(\e)$ the integration can alternatively be performed by expanding the cosine functions as a series obtaining
\begin{align}
B_{a,b}(t,\epsilon') = \frac{b-a}{2}t^2 + \sum_{n=2}^{\infty}  \frac{(b-\epsilon')^{2n-1} - (a-\epsilon')^{2n-1}}{(-1)^{n-1} (2n)! (2n-1)} t^{2n}
\end{align}
for the constant case. 
Assuming the symmetric SIAM by choosing $\e_d = E_d^1 = -U/2$ and 
exploiting the fact that $B_{-D,0}(t,\e) = B_{0,D}(t,-\e)$, we arrive at the final result of Eq. \eqref{eq:exp_value_siam}:
\begin{align}
n_\uparrow(t) &= \frac{1}{2} + B_{0,D}(t,U/2) - B_{0,D}(t,-U/2)\non
n_\downarrow(t) &=  B_{0,D}(t,U/2) +  B_{0,D}(t,-U/2) \\
n_d(t) &= n_\uparrow(t) + n_\downarrow(t) = \frac{1}{2} + 2 B_{0,D}(t,U/2) \non
S_z(t) &= \frac{1}{2} \left( n_\uparrow(t) - n_\downarrow(t) \right) = \frac{1}{4} -  B_{0,D}(t,-U/2) \punkt
\label{eq:analytic_SIAM}
\end{align}
\end{widetext}


\end{document}